\documentclass[12pt, a4paper]{article}

\pdfoutput=1
\usepackage{bbold}
\usepackage{amsmath}
\usepackage{amsfonts}
\usepackage{amssymb}
\usepackage{latexsym}
\usepackage{graphicx}
\usepackage{color}
\usepackage{mathrsfs}
\usepackage{slashed}
\usepackage{soul}
\usepackage{array}
\usepackage{cite}
\usepackage{hyperref}

\usepackage{enumerate}
\usepackage{multirow}
\usepackage[table]{xcolor}
\usepackage{colortbl}

\usepackage{float}

\usepackage{comment}
\includecomment{toexclude} 
\excludecomment{addsections} 

\numberwithin{equation}{section}

 \def\be   {\begin{equation}}   \def\ee   {\end{equation}}
 \def\ba   {\begin{array}}      \def\ea   {\end{array}}
 \def\bea  {\begin{eqnarray}}   \def\eea  {\end{eqnarray}}
 \def\bean {\begin{eqnarray*}}  \def\eean {\end{eqnarray*}}
 \def\ga   {\gamma}
 \def\Ga   {\Gamma}
  \def\la   {\lambda}
  \def\al   {\alpha}
    \def\de   {\delta}
     \def\De   {\Delta}

 \def\nn{\nonumber}
 \def\lee { \left( }
\def\rii { \right) }
\def\lan   {\langle}
\def\ran   {\rangle}
\def\th    {\theta}
\def\ep  {\epsilon}
\def\si {\sigma}
\def\ka {\kappa}

\def\kavu {\kappa\langle \phi_{N_1} \rangle }
\def\kavd {\kappa\langle \phi_{N_2} \rangle }
\def\kavt {\kappa\langle \phi_{N_3} \rangle }

\begin{document}
\hfill\textit{LCTS/2014-23}


\begin{center}
{\Huge
The Dark Side of $\theta_{13}$, $\delta_{CP}$, Leptogenesis\\ [0.9 cm] and Inflation
in Type-I Seesaw
}
\\ [2.5cm]
{\large{\textsc{
Wei-Chih Huang$^{\,a,}$\footnote{\textsl{wei-chih.huang@ucl.ac.uk}} }}}
\\[1cm]

\large{\textit{
$^{a}$~Department of Physics and Astronomy, University College London, UK
}}
\\ [2 cm]
{ \large{\textrm{
Abstract
}}}
\\ [1.5cm]
\end{center}

In the context of the type-I seesaw mechanism, it is known that $\theta_{13}$ is zero and leptogenesis can not be realized if there exists a residual flavor
symmetry resulting in the Tri-Bimaximal neutrino mixing pattern. We propose a simple framework where additional particles, odd under a $Z_2$
symmetry, break the residual flavor symmetry and the lightest of the $Z_2$ odd particles is the dark matter candidate.
As as result, nonzero $\theta_{13}$, $\delta_{CP}$, leptogenesis and the correct dark matter density can be accommodated.
On the other hand, a $Z_2$ odd scalar can play the role of the inflaton with mass of $10^{13}$ GeV motivated by the recent BICEP2 results.
Interestingly, the model can ``generate'' $\delta_{CP} \simeq -\pi/2$, preferred by the T2K experiment in the normal hierarchy neutrino mass spectrum.

\def\thefootnote{\arabic{footnote}}
\setcounter{footnote}{0}
\pagestyle{empty}

\newpage
\pagestyle{plain}
\setcounter{page}{1}

\section{Introduction}\label{sec:Introduction}

Although the evidence for massive neutrinos and the existence of Dark Matter (DM) is well established, neither of them can be explained by the Standard Model (SM). The simplest extension to the SM for providing a mass to neutrinos is to introduce extra SM gauge singlet fermions, coupling to the active neutrinos and the Higgs boson via Yukawa couplings, the so-called type-I seesaw~\cite{Seesaw}. At the same time, any additional SM gauge singlet, with a symmetry to guarantee its stability or having a long lifetime compared to the age of the universe, can be the DM candidate. The main idea of this work is to make a connection between DM and the neutrino sector based on the following observations.

The active neutrino mixing matrix, the Pontecorvo-Maki-Nakagawa-Sakata matrix $U_{PMNS}$, can be well approximated by the Tri-Bimaximal Mixing (TBM) pattern~\cite{Harrison:2002er,Harrison:2002kp,Xing:2002sw,Harrison:2002et,Harrison:2003aw,Harrison:2004he} if 
the third neutrino mixing angle $\th_{13}$ is zero. There have been many models based on discrete flavor symmetries that can naturally create the
TBM pattern. Some of them are $A_4$, for example, Refs.~\cite{ Ma:2001dn, Babu:2002dz, Ma:2005qf, Altarelli:2005yx,King:2006np,Lin:2008aj,Chen:2009um}, $S_4$~\cite{Hagedorn:2006ug,Altarelli:2009gn,Bazzocchi:2009pv,BhupalDev:2011gi,BhupalDev:2012nm}, and $T^\prime$~\cite{Chen:2007afa,Chen:2009gf,Meroni:2012ty}, etc. The idea behind these models is to look for proper group representations for particles in question such that after scalars in the model obtain Vacuum Expectation Values (VEVs) which break the flavor symmetry, the neutrino mixing matrix features the TBM pattern resulting from the residual flavor symmetry. The discovery of non-vanishing $\th_{13}$ in reactor neutrino experiments~\cite{Abe:2011fz,An:2012eh,Ahn:2012nd} (also~\cite{Adamson:2013ue,Abe:2013hdq}), however, demands breaking of the discrete symmetry.

On the other hand, it has been shown in Ref.~\cite{ AristizabalSierra:2009ex} (also Refs.~\cite{Jenkins:2008rb,Bertuzzo:2009im,Hagedorn:2009jy,Felipe:2009rr}) that in the type-I seesaw with non-degenerate neutrino spectra, the $R$ matrix in the Casas-Ibarra parametrization~\cite{Casas:2001sr} is simply a diagonal matrix with elements being $\pm 1$, if there is an underlying discrete flavor symmetry at work. As a consequence, the lepton asymmetry, which is proportional to the imaginary part of $R$, vanishes. 
The underlying reason is that the discrete flavor symmetry usually leads to the form-diagonalizable~\cite{Low:2003dz} neutrino mass matrix, i.e., neutrino masses are completely independent of the mixing matrix elements. In other words, input parameters determining the masses are not related to those determining the mixing angles and phases. The form-diagonalizable property reduces the number of parameters in the rotation matrices used to diagonalize the full neutrino matrix (including both heavy and light neutrinos), leaving $R$ the unit matrix up to a minus sign. Alternatively, another explanation is based on the idea of Form Dominance (FD), that is the requirement that each column of the Dirac mass matrix in the flavor basis is proportional to a different column of the PMNS matrix~\cite{Chen:2009um, Choubey:2010vs}. The type-I seesaw with a flavor symmetry has the FD property.

In this paper, we propose a simple framework, where the underlying flavor symmetry is broken by additional ``dark'' particles, odd under an
imposed $Z_2$ symmetry: an $SU(2)_L$ singlet fermion $\chi_1$, which is the DM candidate, and a fermonic $SU(2)_L$ doublet $\chi_2$ and a real gauge-singlet scalar $S$. The radiative corrections from these particles to the Dirac mass matrix violate the flavor symmetry, leading to nonzero $\th_{13}$ and leptogenesis. The connection between the $CP$-violation phase $\de_{CP}$ in $U_{PMNS}$ and leptogenesis will be established if the only source of $CP$-violation comes from the radiative corrections. We also explore the possibility of $S$ being the inflaton motivated by the recent BICEP2 results on the scalar-to-tensor ratio~\cite{Ade:2014xna}.
All in all, we explore the interplay among $\th_{13}$, $\de_{CP}$, leptogenesis, DM and inflation. Instead of presenting a concrete flavored DM model, we search for a minimum setup with the smallest set of parameters to achieve non-vanishing $\th_{13}$ and leptogenesis, assuming the flavor symmetry yields the TBM pattern. For discussions on flavored DM models, see Refs.~\cite{D'Ambrosio:2002ex,MarchRussell:2009aq,Batell:2011tc,Agrawal:2011ze,Kile:2011mn,Kamenik:2011nb,Batell:2013zwa,Kumar:2013hfa,Lopez-Honorez:2013wla} and the recent review~\cite{Kile:2013ola}. Note that the idea of connecting DM to flavor symmetry breaking (and hence $\th_{13}$ or leptogenesis) has been proposed before, for example, in Refs. \cite{Hambye:2006zn,Choubey:2010vs,Hirsch:2010ru,Boucenna:2011tj,Ahn:2012cga,Ma:2012ez}. This work has two distinctive features. The first is the radiative correction arises in the Dirac mass matrix, instead of the light neutrino mass matrix. Second, our dark matter candidate is {\it{not}} one of right-handed neutrinos, which are even under the $Z_2$ symmetry in this model.

This paper is organized as follows. In Section~\ref{sec:model}, we specify the particle content and quantum numbers.
Section~\ref{sec:theta13_pert} is devoted to the study of the radiative corrections to $\th_{13}$ from the dark particles.  
We discuss the DM relic density in section~\ref{sec:DM} via annihilations through Higgs exchange, and leptogenesis in Section~\ref{sec:lepto} by including new contributions from $\chi$'s and $S$. We present the results in Section~\ref{sec:DM13lepto}, considering the DM density, $\th_{13}$ and leptogenesis. $S$ being the inflaton is discussed in Section~\ref{sec:inflation}. We conclude in Section~\ref{sec:conclusion}.

\section{Model and Observables}\label{sec:model}

The model consists of three heavy right-handed neutrinos, $N_1$, $N_2$ and $N_3$ with $m_{N_3} \geq  m_{N_2} \geq m_{N_1}$. In addition, we have a gauge-singlet fermion $\chi_1$, an fermionic $SU(2)_L$ doublet $\chi_2$, and a real gauge-singlet scalar $S$. Moreover, we impose a $Z_2$ symmetry under which $\chi_1$, $\chi_2$ and $S$ are odd, to guarantee the stability of the DM candidate, $\chi_1$.
The Lagrangian reads\footnote{The two-component spinor notation has been used throughout the paper, unless noted otherwise.}

\be
\mathcal{L} \supset  y_{\al i} \, \lee L_{\al} \cdot H \rii N_{i} - \frac{M_i}{2} N_i N_i + \la_{\al}  \, \lee L_\al \cdot \chi_2 \rii S
+\la_{H\chi } \, \lee \chi_2 \cdot \tilde{H} \rii \chi_1  + \la_{N_i} \, \chi_1 N_i S + h.c. \, ,
\label{eq:lagrangian}
\ee
where  $L_\al=( \nu_\al \,\, e_\al )^T$ and $\al=(e, \mu, \tau)$. $H$ is the SM Higgs doublet, and ``$\cdot$" refers to $SU(2)$ multiplication to form a singlet. An additional $SU(2)_L$ doublet $\tilde{\chi}_2$, with an opposite $U(1)_y$ charge to $\chi_2$, is also introduced to make the model anomaly-free.
We omit mass terms for $\chi_{1,2}$ and $S$, which are not relevant here, and will explicitly specify them when discussing the DM phenomenology.
The quantum numbers of the model are shown in Table~\ref{tab:quantum_number}. Notice that the Lagrangian is invariant under the $SU(2)_L \times U(1)_y$ gauge symmetry but does not necessarily preserve the underlying residual flavor symmetry responsible for the TBM pattern. In fact, we {\it{do}} require residual flavor symmetry breaking to have $\th_{13} \sim 9^\circ$. We provide a simple model based on the $A_4$ symmetry in Appendex~\ref{sec:toyf} to realize
the Lagrangian, Eq.~\ref{eq:lagrangian}.  The goal of this work, however, is to look for the minimum setup to achieve non-vanish $\th_{13}$ without looking into details of the flavor charge assignment.
 
\begin{table}[!h!]
\centering
\begin{tabular}{c c c c c  c c c c c}
  \hline\hline
  Field & $L$ & $H$ & $N_1$ & $N_2$ & $N_3$ &  $\chi_1$ & $\chi_2$& $\tilde{\chi}_2$ & $S$ \\
  \hline
 $SU(2)_L $ &  2 & 2 & 1 & 1& 1 & 1 & 2 & 2 & 1\\
 \hline
  $U(1)_Y $ &  -1/2 & 1/2 & 0 & 0 & 0 & 0 & 1/2& -1/2 & 0\\
 \hline
  $Z_2 $ &  + & + & + & + & + & -- & -- & -- & --\\
  \hline\hline
\end{tabular}
\caption{\emph{The particle content and corresponding quantum numbers in the model. }}
\label{tab:quantum_number}
\end{table}

In this work, we consider the observed $U_{PMNS}$ mixing angles and the light neutrino mass-squared differences, the DM relic density and the baryon density, presented in Table~\ref{tab:observables}. In Section~\ref{sec:theta13_pert}, we fit to only the $U_{PMNS}$ angles and the mass-squared differences in order to show how the existence of $\chi$'s and $S$ can modify $U_{TBM}$ in both the Normal Hierarchy (NH) and Inverted Hierarchy (IH) neutrino mass spectra, while in Section~\ref{sec:DM13lepto} we fit to all observables listed in Table~\ref{tab:observables}.

 \begin{table}[!h!]
\centering
\begin{tabular}{c c c c c  c c c  }
  \hline\hline
   & $\sin^2 2\th_{12}$ &  $\sin^2 2\th_{23}$ & $\sin^2 2\th_{13}$ & $\De m^2_{sol}$ (eV$^2$) &  $|\De m^2_{atm}|$ (eV$^2$) & $\Omega_{\rm{b}} h^2$ & $\Omega_{\rm{DM}} h^2$  \\
  \hline
 best-fit   &  0.857 & 1 & 0.095 &  $7.50\times 10^{-5}$ & $2.32\times 10^{-3}$ & 0.022 & 0.120  \\
 \hline
 $1\si$  &  0.024 & 0.301  & 0.01 & $2\times 10^{-6}$ & $1\times 10^{-4}$  &  $3.3\times 10^{-4}$ & $3.1\times 10^{-3}$ \\
   \hline\hline
\end{tabular}
\caption{\emph{The best-fit value and  $1\si$ standard deviation of relevant observables included in this paper. The values are taken from Refs.~\cite{Beringer:1900zz,Abe:2013fuq,Ade:2013zuv}. }}
\label{tab:observables}
\end{table}

\section{Nonzero $\th_{13}$} \label{sec:theta13_pert}

The idea of this paper is to explore a scenario where the particles odd under the $Z_2$ symmetry break the underlying flavor symmetry\footnote{As we mentioned above, the underlying flavor symmetry has been broken by VEVs of scalars charged under the flavor symmetry. The residual symmetry leads to the TBM pattern. From now on, flavor symmetry refers to ``residual'' flavor symmetry, unless noted otherwise. } and we choose the flavor symmetry to reproduce the TBM pattern\footnote{In general, one can repeat the procedure for arbitrary mixing patterns.} featuring zero $\th_{13}$:
\be
U_{TBM}= \left(
  \begin{array}{c c c}
     \sqrt{ \frac{2}{3} }&  \frac{1}{\sqrt{3}}  & 0 \\
   - \frac{1}{\sqrt{6}} &  \frac{1}{\sqrt{3}}   &   \frac{1}{\sqrt{2}} \\
 -\frac{1}{\sqrt{6}}  & \frac{1}{\sqrt{3}}  &  -\frac{1}{\sqrt{2}}    \\
  \end{array}
\right).
\ee
Then we investigate how the existence of DM can perturb $U_{TBM}$ into $U_{PMNS}$ with $\th_{13} \sim 9^\circ$.
The 6-by-6 neutrino mass matrix is
\be
m= \left(
  \begin{array}{cc}
     0 & m_D \\
    m_D^T & M  \\
  \end{array}
\right),
\ee
where $m_D$ is the Dirac mass matrix and $M$ is the heavy neutrino mass matrix.
As shown in Ref.~\cite{AristizabalSierra:2009ex}, if there is an underlying flavor symmetry at work, $m_D$ can be completely determined by the light and heavy  neutrino masses, $m_{\nu_{1,2,3}}$ and $m_{N_{1,2,3}}$, up to phases, in a basis where $M$~($=\mbox{diag}(m_{N_a},m_{N_b},m_{N_c})$) is diagonal, i.e.,
\be
m_D^0= U_{TBM} P
\left(
  \begin{array}{ccc}
     \sqrt{m_{\nu_1}} & 0 & \\
    0 & \sqrt{m_{\nu_2}} & 0  \\
  0 & 0 & \sqrt{m_{\nu_3}}  \\
  \end{array}
\right)
\left(
  \begin{array}{ccc}
     \sqrt{m_{N_a}} & 0 & \\
    0 & \sqrt{m_{N_b}} & 0  \\
  0 & 0 & \sqrt{m_{N_c}}  \\
  \end{array}
\right).
\ee
Here, $P=\mbox{diag}(e^{i\ga_1},e^{i\ga_2},e^{i\ga_3})$ can, in principle, be absorbed into Majorana phases and is not relevant for this work. Note that we do {\it{not}} assume $m_{N_a} \le m_{N_b} \le m_{N_c}$ and that is why we use $N_{a,b,c}$ instead of $N_{1,2,3}$. 
The superscript 0 on $m_D$ refers to the unperturbed $m_D$ coming from $U_{TBM}$ only.

\begin{figure}
   \centering
 \includegraphics[scale=0.8]{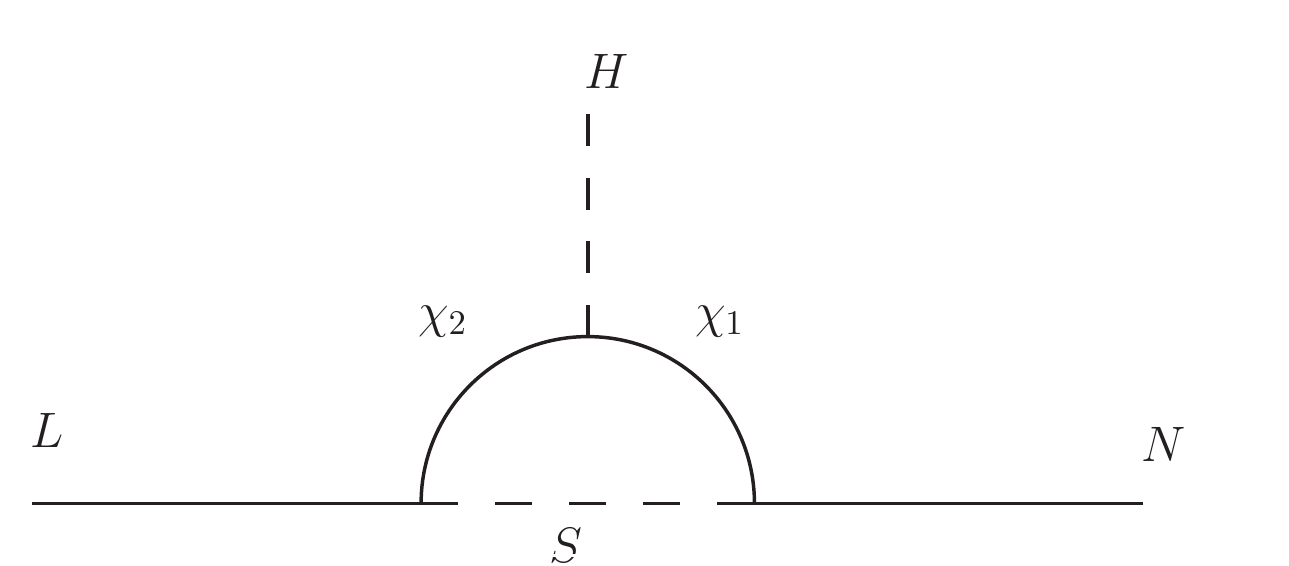}
 \caption{\emph{DM corrections to the Dirac neutrino mass, $m_D$.}}
 \label{fig:L_H_N_loop}
\end{figure}

As shown in Fig.~\ref{fig:L_H_N_loop}, the dark particles running in the loop (DM Loop hereafter) can contribute to $m_D$, which in turn changes the neutrino mixing matrix, and the radiative corrections can be written as,

\be
\de m_D=  \frac{ \langle H^0 \rangle} {\sqrt{2}}  \la_{H\chi} \left(
  \begin{array}{ccc}
     \la_e \la_{N_a} & \la_e \la_{N_b} & \la_e \la_{N_c} \\
     \la_\mu \la_{N_a} & \la_\mu \la_{N_b} & \la_\mu \la_{N_c} \\
      \la_\tau \la_{N_a} & \la_\tau \la_{N_b} & \la_\tau \la_{N_c} \\
  \end{array}
\right) f_{loop} ,
\ee
where $f_{loop}$ is the loop function and $\langle H^0 \rangle=v~(\sim 246)$ GeV is the Higgs VEV. 

In this paper, instead of performing detailed parameter space scans as done in Refs.~\cite{Acosta:2012qf,Acosta:2014dqa,Kile:2014kya}, we keep a spirit of minimality in mind, managing to find a minimal model with the fewest parameters to realize dynamical breaking of the flavor symmetry through the DM loop. Thus, we use the benchmark point in Table~\ref{tab:bechmark_TeV} with $(N_a,N_b,N_c)=(N_1,N_2,N_3)$. In fact, as long as the heavy neutrino masses are of the same order, $\th_{13}\sim 9^\circ$ can always be obtained regardless of the ordering of $m_{N_a}$, $m_{N_b}$ and $m_{N_c}$.  The mass-squared differences, $ m^2_{\nu_2}- m^2_{\nu_1}$ and $|m^2_{\nu_3} - m^2_{\nu_2}|$, are fixed to the observed values $\De m^2_{sol}$ and $\De m^2_{atm}$ from solar and atmospheric neutrino oscillation experiments.\footnote{Note that $m_{\nu_{1,2,3}}$ as input parameters, may {\it{not}} be the same as the resulting light neutrino masses once the DM loop contributions are taken into account. In other words, by fixing the unperturbed $\De m^2$ to be the observed ones, it implies the DM loop can not modify $\De m^2$ significantly.}  With zero $\la_{N_{a(=1)}}$, $\la_{N_{b(=2)}}$ and $\la_\tau$, the DM loop radiative corrections contribute to the $(1,3)$ and $(2,3)$ elements of the Dirac mass matrix $m_D$, denoted as $(\de m_D)_{13}$ and $(\de m_D)_{23}$, respectively.

We should mention that the sum of light neutrino masses, $\sum_i m_{\nu_i}$, is roughly bounded below $0.3$ eV from cosmological constraints from Refs.~\cite{Seljak:2006bg,Joudaki:2012fx,Xia:2012na, RiemerSorensen:2012ve, Zhao:2012xw, Ade:2013zuv,Riemer-Sorensen:2013jsa}. In fact, $m_{\nu_3}=0.1$ eV is in tension with some of the references.

\begin{table}[!h!]
\centering
\begin{tabular}{c c c c c c c}
  \hline
  \hline
   & $m_{\nu_1}$ (eV) &  $ m_{\nu_2}$ (eV) &   $  m_{\nu_3}$ (eV) 
    & $\la_{N_a}$ & $\la_{N_b}$ &$\la_{\tau}$ \\
  \hline 
 NH  &  0  & $8.66\times 10^{-3}$  & $4.89\times 10^{-2}$  & 0 &0 & 0\\
 \hline 
   IH &  $1.107 \times 10^{-1}$ & $1.11\times 10^{-1}$   & 0.1  & 0 &0 & 0 \\
  \hline\hline
     & $m_{N_1}$ (GeV) & $m_{N_2}$ (GeV) & $m_{N_3}$ (GeV) & $m_{S}$ (GeV) & $m_{\chi_1}$ (GeV) & $m_{\chi_2}$ (GeV) \\
  \hline
 NH/IH  & 1000 & $1000+\De m_{N_{12}}$ & 2000 & 700 & 62 & 200  \\
  \hline
\end{tabular}
  
\caption{\emph{The benchmark point for the NH and IH cases. We here keep $ m^2_{\nu_2}- m^2_{\nu_1}$ and $|m^2_{\nu_3} - m^2_{\nu_2}|$ to be $\De m^2_{sol}$ and $\De m^2_{atm}$, respectively. The reason why $m_{\chi_1} \sim m_h/2$ comes from the resonant enhancement from the DM consideration as we shall see below. In addition, we also need $\De m_{N_{12}}\equiv m_{N_2}-m_{N_1}\sim \Ga_{N_2}$~(decay width of $N_2$) for the resonant enhancement to realize  low-scale leptogenesis.}} 
\label{tab:bechmark_TeV}
\end{table}

\begin{figure}[!htb!]
   \begin{minipage}{0.4\textwidth}
   \centering
   \includegraphics[scale=0.8]{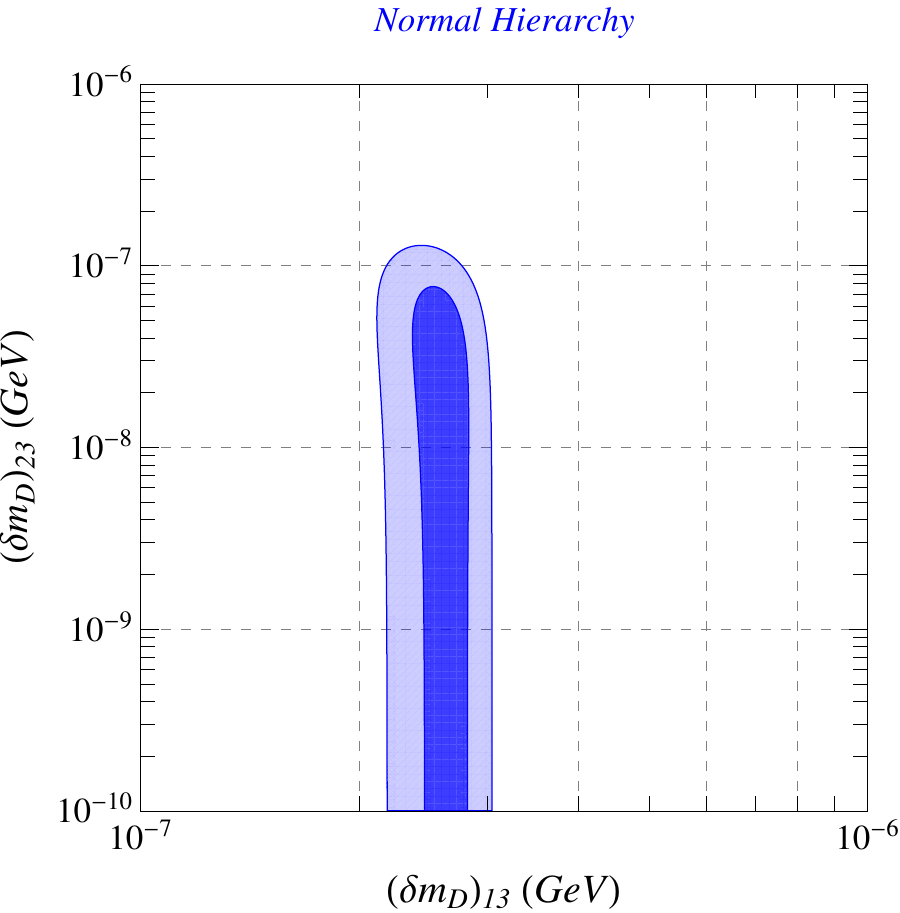}
    \end{minipage}\hspace{2 cm}
   \begin{minipage}{0.4\textwidth}
    \centering
    \includegraphics[scale=0.8]{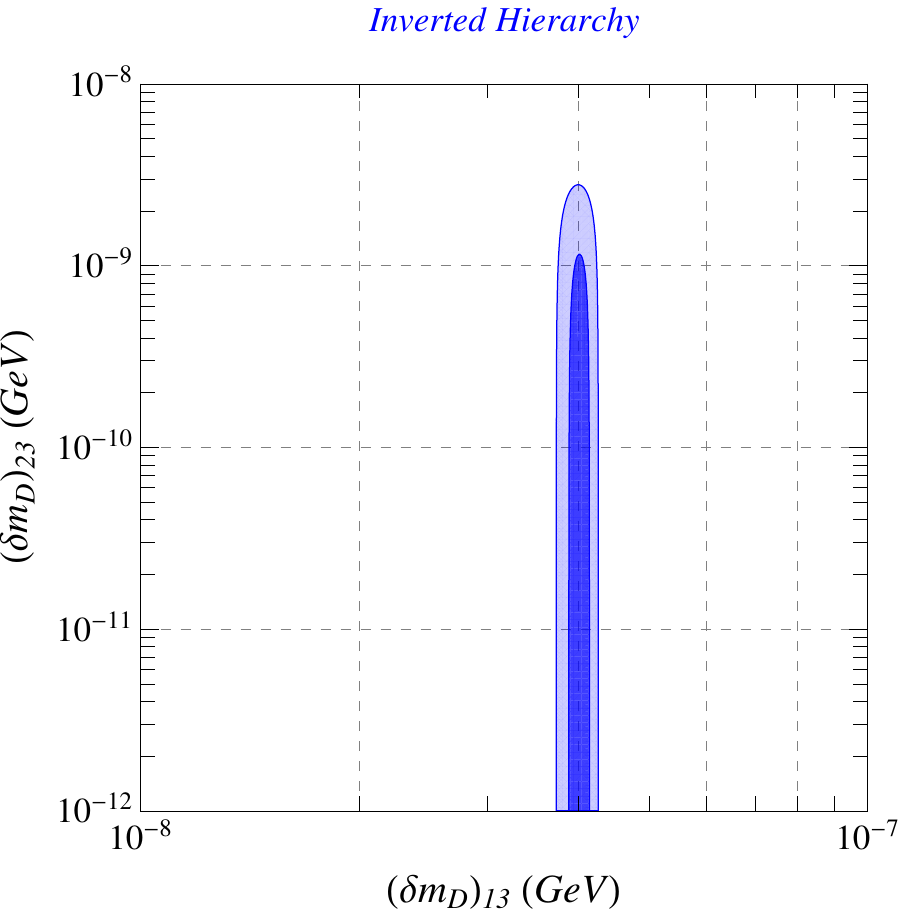}
    \end{minipage}
\caption{\emph{ Confidence region on $\lee \de m_D \rii_{13}$ and $\lee \de m_D \rii_{23}$ to reproduce the observed neutrino mixing angles and the mass-squared differences via the DM loops using the benchmark point in Table~\ref{tab:bechmark_TeV} with $(N_a,N_b,N_c)=(N_1,N_2,N_3)$. For the NH case, the best reduced $\chi^2$, $\chi^2$ per degree of freedom, is 0.46 while the reduced $\chi^2$ is 1.29 for IH. }}
\label{fig:le_lmu}
\end{figure}

For the fitting procedure, we first compute the modified light neutrino mass spectrum and the mixing angles as the functions of $\la_e$ and $\la_\mu$ with  the benchmark point in Table~\ref{tab:bechmark_TeV} and $\la_{N_c}=1$.
Then, we fit to the observed $U_{PMNS}$ angles and the mass-squared differences shown in Table~\ref{tab:observables}.
The results are shown in Fig.~\ref{fig:le_lmu}, with $68\%$ (dark blue) and $99\%$ (light blue) confidence region. Note that given the neutrino mass matrix $m$, there are ambiguities on determining the mixing angles of $U_{PMNS}$. To be more concrete, any equivalent transformation, $U^{\prime}_{PMNS} \rightarrow P_1 U_{PMNS} P_2$ in which $P_{1,2}$ are diagonal matrices with elements of $\pm 1$, renders intact the physical observables, such as oscillation probabilities, but will correspond to different active mixing angles. As demonstrated in Refs.~\cite{deGouvea:2008nm,deGouvea:2010iv}, all mixing angles can be chosen positive and smaller than or equal to $\pi/2$ provided $\de_{CP}$ is allowed to vary between $-\pi$ and $\pi$. For this reason, we choose to use $\sin^2 2\th$'s in the fit, which are free from ambiguities.

Remarkably, $\la_e$ (and $\la_{N_c}$) alone can amend $m_D$ to produce desired $U_{PMNS}$ and $\De m^2$, for both IH and NH.\footnote{For the IH case, the reason why $m_{\nu_3}$ has to be nonzero is we constrain ourselves to real $\la_e$ and $\la_{\mu}$. With complex $\la$'s, zero $m_{\nu_3}$ can be achieved.}
In the NH situation, this can be understood by simply looking at the perturbed $m_D$ (with $m_{\nu_1}$=0) including $\la_e$ and $\la_{N_c}$ only,
\be
m_D= m_D^0 + \de m_D=\left(
  \begin{array}{ccc}
   0 & \sqrt{ \frac{ m_{N_b}  m_{\nu_2} }{3 }} & \lee \de m_D \rii_{13}  \\
  0 & \sqrt{ \frac{ m_{N_b}  m_{\nu_2} }{3} } &  - \sqrt{ \frac{ m_{N_c}  m_{\nu_3} }{2} } \\
   0 & \sqrt{ \frac{ m_{N_b}  m_{\nu_2} }{3} } & \sqrt{ \frac{ m_{N_c}  m_{\nu_3} }{2} } \\
  \end{array}
\right),
\ee
and the light neutrino mass
matrix is
\bea
m_\nu  \hspace{-5mm} && \sim  - m_D M^{-1} m_D^T  \\
&& = - \left(
  \begin{array}{ccc}
    \frac{\lee \de m_D \rii_{13}^2}{m_{N_c}}+ \frac{ m_{\nu_2}}{3 } &  - \lee \de m_D \rii_{13} \sqrt{ \frac{ m_{\nu_3} }{2 m_{N_c}}} + \frac{ m_{\nu_2}}{3 }
    &  \lee \de m_D \rii_{13} \sqrt{ \frac{ m_{\nu_3} }{2 m_{N_c}}} + \frac{ m_{\nu_2}}{3 }  \\
  - \lee \de m_D \rii_{13} \sqrt{ \frac{ m_{\nu_3} }{2 m_{N_c}}} + \frac{ m_{\nu_2}}{3 }  & \frac{m_{\nu_2}}{3} + \frac{m_{\nu_3}}{2}
  &  \frac{m_{\nu_2}}{3} - \frac{m_{\nu_3}}{2}  \\
     \lee \de m_D \rii_{13} \sqrt{ \frac{ m_{\nu_3} }{2 m_{N_c}}} + \frac{ m_{\nu_2}}{3 } & \frac{m_{\nu_2}}{3} - \frac{m_{\nu_3}}{2}
     &  \frac{m_{\nu_2}}{3} + \frac{m_{\nu_3}}{2}  \nn \\
       \end{array}
\right),
\label{eq:perturbed_13}
\eea
where $\lee \de m_D \rii_{13}$ denotes the DM loop contribution.
The existence of $\lee \de m_D \rii_{13}$ explicitly breaks the residual $\mu-\tau$ symmetry, making $\th_{13} \neq 0$~\cite{Ma:2002ce}.
In addition, we have $\lee \de m_D \rii_{13}\sim 10^{-7}$ GeV from the confidence region in Fig.~\ref{fig:le_lmu}, and in turn $\frac{\lee \de m_D \rii_{13}^2}{m_{N_c}}  \ll  \frac{m_{\nu_2}}{3}$
so that the trace of $m_\nu$, the sum of three light neutrino masses, remains unchanged, i.e., the mass-squared differences
stay intact. The reason why $\lee \de m_D \rii_{13}$ is so small is that the neutrino mass-squared differences are fixed to the experimental values in the benchmark points so that the radiative correction is forced to be small in order to reproduce the neutrino oscillation observables. 
In summary, the DM loop with $\la_e$ (and $\la_{N_c}$) induces the $\nu_e-N_c$ mixing which then breaks the $\mu-\tau$ symmetry to generate sizable $\th_{13}$ but keep the light neutrino mass spectrum unscathed. We refer readers to Ref.~\cite{Liao:2012xm}, where different breaking patterns on the $\mu-\tau$ symmetry have been studied systematically, and also Refs.~\cite{Xing:2002sw,Hirsch:2003dr} on modifications or radiative corrections to the TBM pattern.   Furthermore, $(\de m_D)_{13}$ is the only radiative correction that can change $U_{TBM}$ into $U_{PMNS}$ and have the correct mass-squared
differences on its own. One must need at least two radiative corrections to achieve the goals if $(\de m_D)_{13}$ is not involved\footnote{This conclusion might change if input parameters, $\De m^2_{12}$ and $\De m^2_{23}$, are allowed to vary.}.  

It is worthwhile to mention that the DM loop, in addition to a real component, would have an imaginary part if the internal particles are on-shell. One might naively conclude it could contribute to a $CP$-violating phase in $U_{PMNS}$. It, however, is a false statement since first the corresponding antiparticles would have the identical loop structure with the same imaginary part due to $CPT$ invariance, i.e., the imaginary part gives rise to a $CP$-conserving phase. Second, this type of the $CP$-conserving phase should not be taken into account when diagonalizing the neutrino mass matrix but should be absorbed into the decay width of the heavy neutrinos. On the other hand, the $CP$-conserving phase does play a role in the context of leptogenesis as discussed below.

\begin{figure}[!htb!]
   \begin{minipage}{0.4\textwidth}
   \centering
   \includegraphics[scale=0.8]{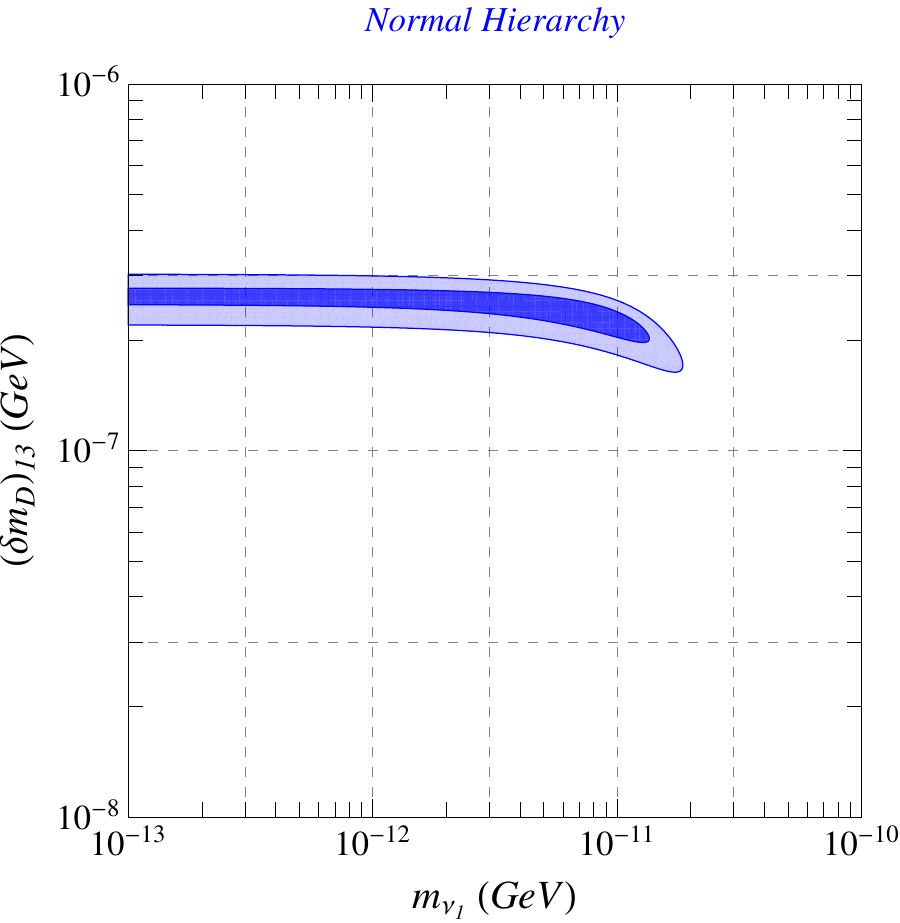}
    \end{minipage}\hspace{2 cm}
   \begin{minipage}{0.4\textwidth}
    \centering
    \includegraphics[scale=0.8]{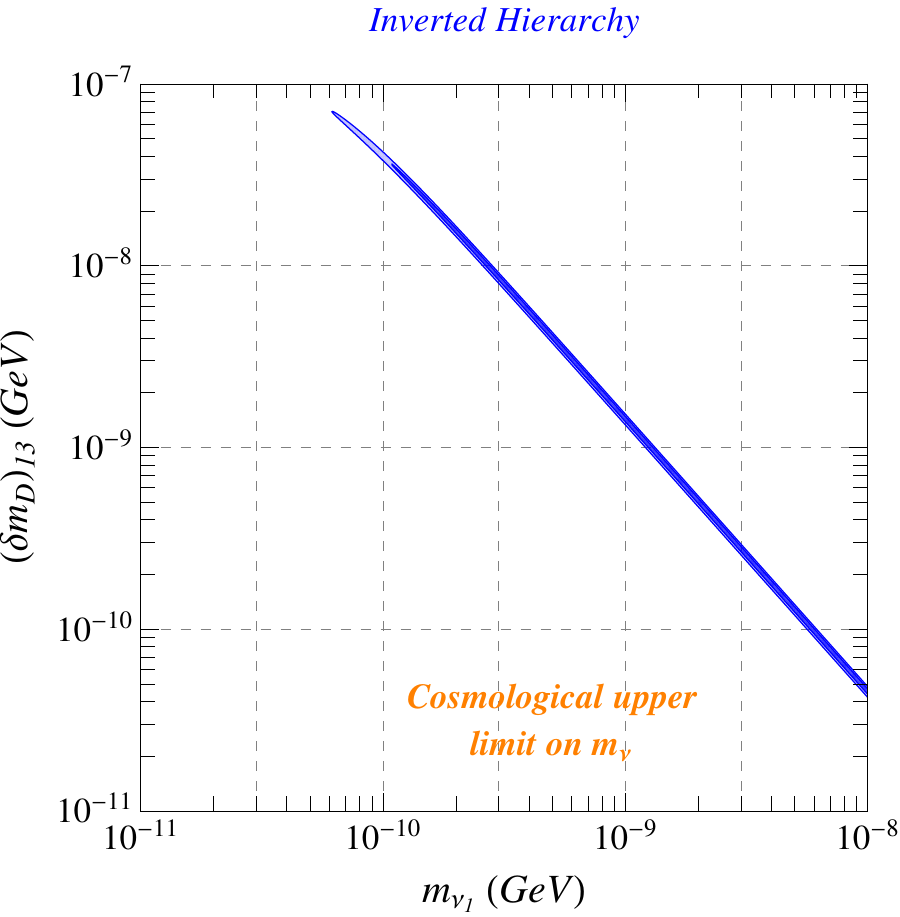}
    \end{minipage}
\caption{\emph{ Confidence region on $m_{\nu_{1,3}}$ and $\lee \de m_D \rii_{13}$ to reproduce the observed neutrino mixing angles and the mass-squared differences via the DM loops using the benchmark point shown in Table~\ref{tab:bechmark_TeV} with $(N_a,N_b,N_c)=(N_1,N_2,N_3)$. Note that only a single DM radiative correction, $\lee \de m_D \rii_{13}$, is included in the fit.
For the NH case, the best reduced $\chi^2$ is 0.25, while the reduced $\chi^2$ is 0.37 for IH. We also show the cosmological constraint~\cite{Seljak:2006bg,Joudaki:2012fx,Xia:2012na, RiemerSorensen:2012ve, Zhao:2012xw, Ade:2013zuv,Riemer-Sorensen:2013jsa} on the neutrino mass. }}
\label{fig:le_mnu1}
\end{figure}

We conclude this section with Fig.~\ref{fig:le_mnu1}, where the DM loop involves $\la_e$ only.
 $m_{\nu_1}$ ($m_{\nu_3}$) for NH (IH) and $\la_e$ are being varied to find the minimum $\chi^2$, while the same set of mass parameters as above
 are assumed. It is clear that in the NH case, small $m_{\nu_1}$ is preferred and $\la_e$ is nearly constant in the confidence region  because of negligible contributions from nearly zero $m_{\nu_1}$. On the other hand, for the IH case, large $m_{\nu_3}~(\gtrsim 0.1$ eV) is preferred, which is compensated by small $\la_e$. The different behaviors can be simply understood by looking into the Dirac mass matrix, $m_D$. In the NH case, $\de m_D$ from the DM loop and terms involving $m_{\nu_1}$ are located in the third and first column, respectively, while for IH, they both appear in the third column. As a consequence, when solving for the light neutrino masses and $U_{PMNS}$, the magnitude of $m_{\nu_3}$ is correlated with $\la_e$ in IH whereas $\la_e$ is nearly independent of $m_{\nu_1}$ in NH.

\section{DM Relic Density} \label{sec:DM}
\begin{figure}
   \centering
 \includegraphics[scale=0.8]{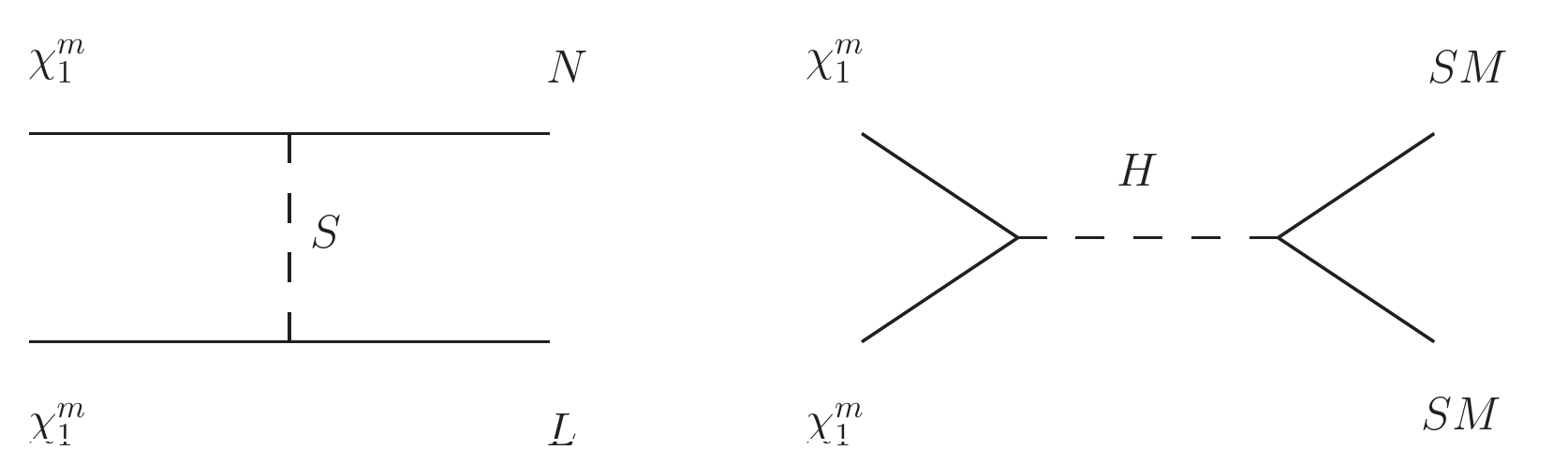}
 \caption{\emph{ Annihilation processes of $\chi_1^m$, where $m$ refers to the mass eigenstate.}}
 \label{fig:coannihilation}
\end{figure}

In this Section,  we compute the DM relic abundance through $\chi_1$ annihilations.
We begin with the relevant Lagrangian  for the DM relic density,
\bea
\mathcal{L}  \supset   && \la_{H\chi } \, \lee \chi_2 \cdot \tilde{H} \rii \chi_1 + \la_{H \tilde{\chi} } \, \lee \tilde{\chi}_2 \cdot H \rii \chi_1 
 + \la_{\al}  \, \lee L_\al \cdot \chi_2 \rii S \nn\\
 && + \la_{N_i}\, \chi_1 N_i S 
  -\frac{1}{2} m_S^2
- \frac{1}{2} m_{\chi_1} \chi_1\chi_1 - m_{\chi_2} \tilde{\chi}_2\chi_2 + h.c.,
\eea
where $H=\frac{1}{\sqrt{2}}( 0 \, , \, v+ h )^T$, in which $h$ is the Higgs boson field.
After  electroweak symmetry breaking, $\chi_1$ generally mixes with the neutral components of $\chi_2$ and $\tilde{\chi_2}$, referred as $\chi^0_2$ and $\tilde{\chi}^0_2$, respectively; therefore DM is the linear combination of $\chi_1$, $\tilde{\chi}^0_2$ and $\chi^0_2$. The DM relic abundance is determined by the processes shown in Fig.~\ref{fig:coannihilation}, where $m$ denotes the mass eigenstate.
The first process is, however, kinematically suppressed since $m_{N} \gtrsim m_{\chi_1}+m_{\chi_2}$ due to the leptogenesis consideration as we shall see below. For the Higgs exchange process, for simplicity, we assume $m_{\chi_2} > m_{\chi_1}$ so that co-annihilation processes are negligible.

In this framework, Direct Detection (DD) constraints, especially those set by LUX~\cite{Akerib:2013tjd} on Spin-Independent (SI) interactions, should be considered since $\chi^m_1$ can also interact with nucleons via $t$-channel Higg-mediated processes. We show the annihilation cross-section and the SI DM-nucleon cross-section in Appendix~\ref{sec:DMDD}. It turns out the required $\la_{H\chi}$'s in order to produce the correct DM density will also generate a large DM-nucleon cross-section, in conflict with the LUX results. There are at least two solutions -- resonant enhancement and co-annihilation. 

\begin{figure}
   \centering
 \includegraphics[scale=0.8]{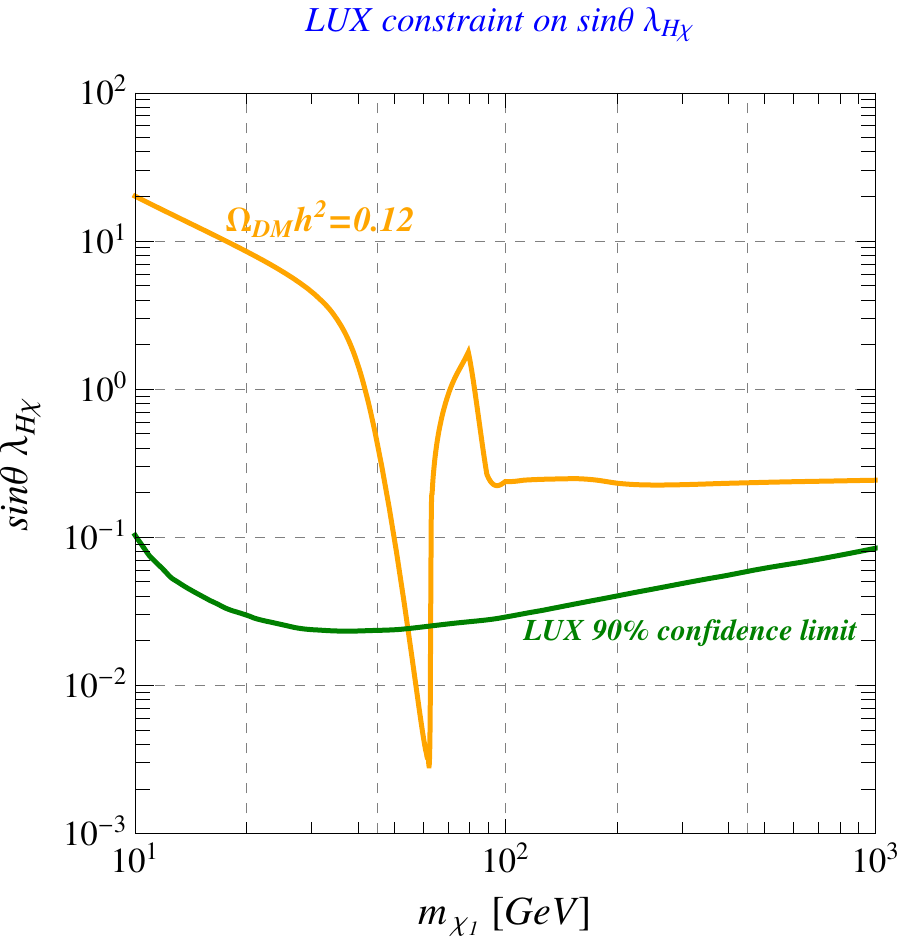}
 \caption{\emph{LUX bounds on the product of $\sin\th$ and $\la_{H}$, where $\th$ is the $\chi_1 - \chi_2$ mixing angle. The Orange line corresponds to the correct DM density, where only annihilation via Higgs exchange is taken into account. The green line is the LUX  $90\%$ confidence limit on SI interactions\cite{Akerib:2013tjd}. It is clear only the resonant region is not excluded by the LUX results; therefore we choose $m_{\chi_1}=62$ GeV as our benchmark point.}}
 \label{fig:LUX}
\end{figure}

\begin{enumerate}
\item One can make $m_{\chi_1} \sim m_h/2$ to enhance the annihilation cross-section by virtue of the small Higgs decay width~($\sim 4$ MeV~\cite{Djouadi:1997yw}) to keep $\la_{H\chi}$ small enough not to be excluded by the LUX DD bounds. From Fig.~\ref{fig:LUX}, we show the LUX constraints on $\la_{H\chi}$ and the $\chi_1-\chi_2$ mixing, $\th$. It is clear that only when $m_{\chi_1}\sim m_h/2$, can $\chi_1$ annihilation be sufficient enough to have the correct density without inducing the large SI DM-nucleon cross-section, avoiding the LUX bounds. As a consequence, we pick $m_{\chi_1}=62$ GeV as our benchmark point.\footnote{For the resonant region, $\chi_1$ mostly annihilates into $b$-quarks, which in turn produce protons and antiprotons in addition to  gamma rays. Stringent limits on the $b$-quark final state is recently derived in Ref.~\cite{Tavakoli:2013zva}, based on indirect DM searches. We would like to point out these limits become much weaker in our model since the annihilation cross-section is velocity suppressed and the current DM velocity is very small~$(\sim 10^{-3})$. }

\item The second method is to make $m_{\chi_2}   \gtrsim  m_{\chi_1}$ to turn on co-annihilation processes such that the DM density is mostly determined by co-annihilation which is not constrained by DD, as long as the mass-splitting is much larger than the typical nuclear recoil energy of order $\mathcal{O}$(KeV) in DD experiments. This solution is quite fine-tuned since, in this setup, annihilation and co-annihilation cross-section are generally of the same order.   

\end{enumerate}

In principle, one should also take into account processes mediated by the $Z$ boson.
For DM annihilation, we simply chose $\la_{H\chi} \sim \la_{H \tilde{\chi}}$ and $m_{\chi_2} \sim m_{\tilde{\chi}_2}$ so that after diagonalizing the mass matrix of $\chi$'s, the DM particle has roughly equal $\chi_2$ and $\tilde{\chi}_2$ component, and consequently does not strongly interact with $Z$ since  $\chi_2$ and $\tilde{\chi}_2$ carry opposite charges. As a result, the Higgs exchange processes mentioned above will be the dominant contribution.
In terms of DD experiments, because of the negligible mixing between $\chi_1$ and $\chi_2$'s, the DM particle is mostly $\chi_1$, that is a Majorana particle. It has only Spin-Dependent (SD) interactions with nucleons through vector boson exchange. The DD bounds on SD interactions are much weaker and thus will not be considered here.

\section{Leptogenesis with TeV $N_1$} \label{sec:lepto}
\begin{figure}
   \centering
 \includegraphics[scale=0.8]{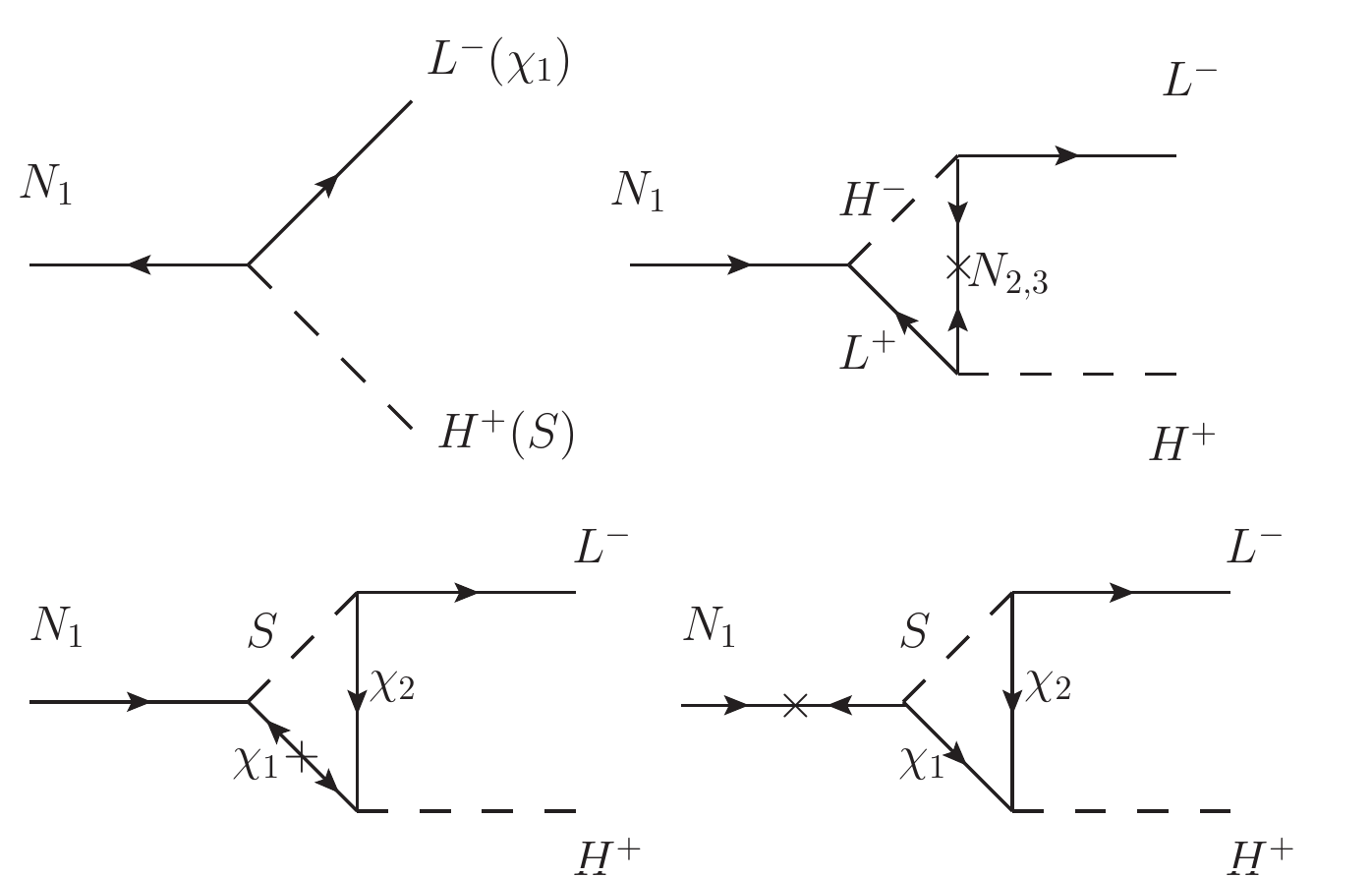}
 \caption{\emph{Processes relevant for leptogenesis including additional contributions from $\chi$'s and $S$.}}
 \label{fig:leptogenesis}
\end{figure}

In this section, we study the lepton asymmetry generated from $N_1$ decays, including additional  contributions from the dark particles $\chi$'s and $S$.
The heavy neutrino mass in question is of order TeV, as shown in Table~\ref{tab:bechmark_TeV}.
Fig.~\ref{fig:leptogenesis} shows the relevant Feynman diagrams for leptogenesis and here we only show $N \rightarrow H^+ L^-$ for demonstration.\footnote{
The arrow on fermion lines represents the chirality of particles and we use the convention in Ref.~\cite{Dreiner:2008tw}, i.e., a particle with the arrow pointing in the same direction as the four momentum is left-handed. We refer readers to the reference for the details of two-component Weyl-spinor notations and  computation techniques. Here, we show only the vertex contribution (top right panel) as a representative of type-I seesaw loop diagrams. In fact, the wave function contribution is also included in computation.}

If there are no DM loop contributions, because of the flavor symmetry
the $R$ matrix is real, leading to zero lepton asymmetry from $N_1$ decays. Additionally, even if $R$ is complex, the generated lepton asymmetry is still too small to account for the observed baryon asymmetry as shown in Ref.~\cite{Davidson:2002qv}: $m_{N_1}$ has to be larger than $10^9$ GeV in order for leptogenesis to work in the type-I seesaw. 

On the other hand, the DM loop can interfere with the tree-level diagram~($N \rightarrow H L$) to induce the lepton asymmetry.
Besides, the $R$ matrix, which was real in the presence of the exact
flavor symmetry, can become complex due to phases in the DM loop and the lepton asymmetry can be generated via conventional leptogenesis as shown in the top panels of Fig.~\ref{fig:leptogenesis}. Nonetheless, it is usually suppressed compared to the one coming from the
direct interference between $N \rightarrow H L$ and the DM loops, the lower panels of Fig.~\ref{fig:leptogenesis}, because of loop suppression and
smallness of Yukawa couplings, $y_{\al i}\sim 10^{-6}$ for $m_{N_i}\sim$ TeV.

Before computing the lepton
asymmetry, one has to remember that in order to obtain nonzero results from tree- and loop-level interference, two conditions must be satisfied -- internal particles in the loop should be
on-shell and the product of couplings of the tree- and loop-diagram must have a nonzero imaginary part. Failing to satisfy either results in zero lepton asymmetry.
As long as $N_1$ is heavier than the sum of $\chi_1$ and $S$ mass, which is the case in this work, the first condition holds. To elaborate the second one,
we have to look at the relevant terms involved in the loop:
\be
\mathcal{L}  \supset  \la_{\al} \lee L_\al \cdot \chi_2 \rii S+  \la_{H\chi } \, \lee \chi_2 \cdot \tilde{H} \rii \chi_1   + \la_{N_1}\, \chi_1 N_1 S.
\ee
Assuming all couplings are complex to begin with: $ \la_{\al} =  | \la_{\al}  | e^{i\th_{\al}} $, $ \la_{H\chi} =  | \la_{H\chi}  | e^{i\th_{H}} $
and $ \la_{N_1} =  | \la_{N_1}  | e^{i\th_1} $, we can make $ \la_{H\chi}$ real by redefining $\chi_2 \rightarrow \chi_2 e^{i\th_H}$.
Consequently, we can redefine $\tilde{\chi}_2 \rightarrow \tilde{\chi}_2 e^{ - i \th_H}$, such that $e^{i\th_{H}}$ is completely removed from the Lagrangian. However, the trick does not work for $\th_1$ and $\th_{\al}$ because ($i$) $\chi_1$ and $N_1$ have Majorana mass terms, i.e., the absorbed phase will show up in the mass terms\footnote{The real scalar $S$ has the same property as well.}, unlike $\chi_2$ and $\tilde{\chi}_2$ which have a Dirac mass term, and ($ii$) removing $\th_\al$ by $L_\al$ redefinition will make Yukawa couplings, $y_{\al i}$, contain $\th_\al$. As we shall see below, radiative corrections to leptogenesis from the dark sector always involve the product of $\la_\al$ and $\la_{N_1}$. 
To simplify the analysis, we make $\la_\al$ real and then the phase of $\la_{N_1}$ will be the only source of the imaginary part.

The lepton asymmetry on flavor $\al$ characterized by $\ep_{\al\al}$ is:
\be
\ep_{\al\al}= \frac{ \Ga(N_1 \rightarrow \ell_\al^{-} H^{+}) - \Ga(N_1 \rightarrow \ell_\al^{+} H^{-})  }
{  \Ga(N_1 \rightarrow \ell_\al^{-} H^{+}) + \Ga(N_1 \rightarrow \ell_\al^{+} H^{-})  + \Ga(N_1 \rightarrow\chi_1 S) },
\ee
where we have neglected the loop contribution in the denominator which is subdominant to the tree-level.
We also consider the dilution from $N_1 \rightarrow \chi_1  S$,
which will not generate a lepton asymmetry. Results of $\ep_{\al\al}$'s from different contributions are shown in Appendix~\ref{sec:lep_com}.
We would like to make two comments about the structure of the loops in Fig.~\ref{fig:leptogenesis}.
 
\begin{itemize}
\item One of the DM loops does not have a mass insertion to flip the chirality of $\chi_1$. In other words, for the bottom-right diagram, lepton number violation required for leptogenesis actually comes from the external $N_1$  unlike the other loop diagrams in Fig.~\ref{fig:leptogenesis}. In the limit of $m_{N_1}>>m_{\chi_1}$, it would be the dominant contribution.

\item  As mentioned before, particles in the loop have to be on-shell. Consequently, the sum of $m_{\chi_1}$ and $m_S$ has to be smaller than $m_{N_1}$ as seen easily from terms like $B_0(m_{N_1}^2, m_S^2, m_{\chi_1}^2)$  in $\ep_{\al\al}$, where $B_0$ is the Passarino-Veltman Integral~\cite{Passarino:1978jh}.

\end{itemize}

 In terms of $\ep_{\al\al}$, the generated lepton asymmetry $Y_{\De L}$ through $N_1$ decays and the resulting baryon asymmetry $Y_{\De B}$
 can be expressed as~\cite{Davidson:2008bu}
 \be
 Y_{\De B}=C Y_{\De L} =C \frac{135 \zeta(3)}{4 \pi^4 g_{*}} \sum_{\al} \ep_{\al\al} \eta_\al,
 \label{eq:Y_B}
 \ee
where $g_{*}$ is the relativistic degrees of freedom when $N_1$ decays.\footnote{In our case, it is 106.75 (from the SM particles)$+$ 1 (from $S$) +  $\frac{7}{8}\times2$ (from $\chi_1$)+ $\frac{7}{8}\times4$ (from $\chi_2$ and $\tilde{\chi}_2$).}
$\eta_{\al}$ characterizes the wash-out effect,\footnote{We follow methods used in Ref.~\cite{Davidson:2008bu} to estimate $\eta_\al$.} and $C$~($= - 28/79 $)~\cite{Khlebnikov:1988sr, Harvey:1990qw} comes from the conversion of $\De L$ into $\De B$ by the sphaleron~\cite{Klinkhamer:1984di, Arnold:1987mh,  Arnold:1987zg}. For recent reviews on leptogenesis, see, for example, Refs.~\cite{Buchmuller:2005eh, Chen:2007fv, Davidson:2008bu}.

\section{DM, $\th_{13}$ and Leptogenesis} \label{sec:DM13lepto}
 Combining all ingredients discussed above, one can check if the DM loop with $\la_e$ and $\la_{c(=1)}$ only can alter $U_{TBM}$ into $U_{PMNS}$, and concurrently accommodate leptogenesis and DM. The difficulty, however, arises from leptogenesis consideration.
 If only the DM loop contribution is considered, the zero tree-level $N_1 \rightarrow L^{\mp}_e H^{\pm}$ (from $(U_{TBM})_{13}=0$) yields a vanishing lepton asymmetry, recalling that the lepton asymmetry requires both the tree- and loop-level contribution. On top of that, the original leptogenesis (top panels of Fig.~\ref{fig:leptogenesis}) does not work either on account of small Yukawa couplings unless the resonance enhancement is involved as we shall see below. To circumvent the problem, one could in principle involve more parameters in the game. With the spirit of minimality in mind, we have two simplest options as follows. 
 
 \begin{enumerate}
 
\item With the additional $\la_\mu$, one might expect that  $(\de m_D)_{13}$, from $\la_e$ (and $\la_{N_1}$), can lead to sizable $\th_{13}$, while $(\de m_D)_{23}$, from $\la_\mu$ (and $\la_{N_1}$), is responsible for leptogenesis. From Eq.~\ref{eq:Y_B} and those in Appendix~\ref{sec:lep_com}, one can infer, in the limit of $m_{N_1} \gg m_{\chi}, \, m_{S}$,
\be
\lee \frac{Y_{\De B}}{ 10^{-10}} \rii \sim \lee \frac{ m_{\chi_1} }{  m_{N_1} } \rii  \lee \frac{ \la_{H\chi} \mbox{Im} \lee \la_{N_1} \rii \la_e}{10^{-10}} \rii
\lee \frac{ 10^{-6} }{ y_{\al i} } \rii.
\ee
With $\la_{H\chi} \sim 0.1$ from the relic density consideration, successful leptogenesis requires $|\la_e \mbox{Im} \lee \la_{N_1}\rii|\sim 10^{-9}$ for TeV $N$'s and sub-TeV $m_\chi$ and $m_S$. On the other hand, for the NH case, generating sizable $\th_{13}$ demands\footnote{It can be derived by simply equating Eq.~\ref{eq:perturbed_13} with $\lee U^*_{PMNS}\cdot \mbox{diag}(m_{\nu_1},m_{\nu_2},m_{\nu_3}).U^\dagger_{PMNS} \rii$, and taking the limit of $m_{\nu_{1,2}} \ll m_{\nu_3}$. The same approach can be applied to the IH case, with $m_{\nu_1} \sim m_{\nu_2} \sim m_{\nu_3}$ for $m_{\nu_3}=0.1$ eV.}
\be
\lee  \de m_D  \rii_{13} \sim  \frac{1}{16 \pi^2}  \lee \la_{H\chi}  \la_{N_1} \la_\mu \rii \frac{v}{\sqrt{2}}
\sim \sqrt{ 2 m_{N_1}  m_{\nu_3}  }  \sin\th_{13}.
\ee
It implies $|\la_\mu \la_{N_1}| \sim 10^{-4}$. In order not to dilute the lepton asymmetry from $N_1$ decaying into $S$ and $\chi_1$ , $\la_{N_1}$ has to be much smaller than Yukawa couplings~($\sim 10^{-6}$). This implies $\la_\mu\gtrsim 10$, and perturbativity is lost.

\item One can also include $\la_{N_2}$ to the model besides $\la_e$ and $\la_{N_1}$.
$\la_{N_1}$  and $\la_e$  are used to accomplish leptogenesis while $\la_{N_2}$  and $\la_e$ give rise to required perturbation on $U_{TBM}$.\footnote{Alternatively, $\th_{13}$ can arise from $N_1$ ($\la_{N_1}$) while leptogenesis comes from $N_2$ ($\la_{N_2}$). Although the generated lepton
asymmetry from $N_2$ will in principle be washed out by $N_1$ decays, there exist situations~\cite{Barbieri:1999ma,DiBari:2005st,Vives:2005ra,Blanchet:2006dq,Strumia:2006qk,Engelhard:2006yg} where the asymmetry survives from $N_1$ washout effects.} This option, however, suffers from new washout effects from $\chi_1 + S \leftrightarrow H^{\pm} + L^{\mp}$ since Majorana $\chi_1$ can not carry the lepton number. For $m_{N_1} \lesssim 10^7$ GeV~\cite{Sierra:2013kba,Racker:2013lua}, washout effects are generally faster than the expansion of the universe, erasing the lepton asymmetry from $N_1$ decays. 

\end{enumerate}
Alternatively, one can employ the resonant enhancement to achieve low-scale leptogenesis~\cite{Pilaftsis:1997jf,Pilaftsis:2003gt}, i.e.,
$m_{N_2}-m_{N_1}\sim \Ga_{N_2}$ ($N_2$ decay width). We here adopt this method to perform $\chi^2$ fits.
To sum up, with a single DM loop associated with $\lee \de m_D \rii_{13}$, the $R$ matrix becomes complex and low-scale leptogenesis
can be realized via the resonant enhancement such that the sufficient lepton asymmetry survives from the washout effects. In this situation, the only $CP$-violation source comes from the phase of $\la_{N_1}$, inducing a connection between $\de_{CP}$ in $U_{PMNS}$ and leptogenesis. 

\begin{figure}[!htb!]
   \begin{minipage}{0.4\textwidth}
   \centering
   \includegraphics[scale=0.8]{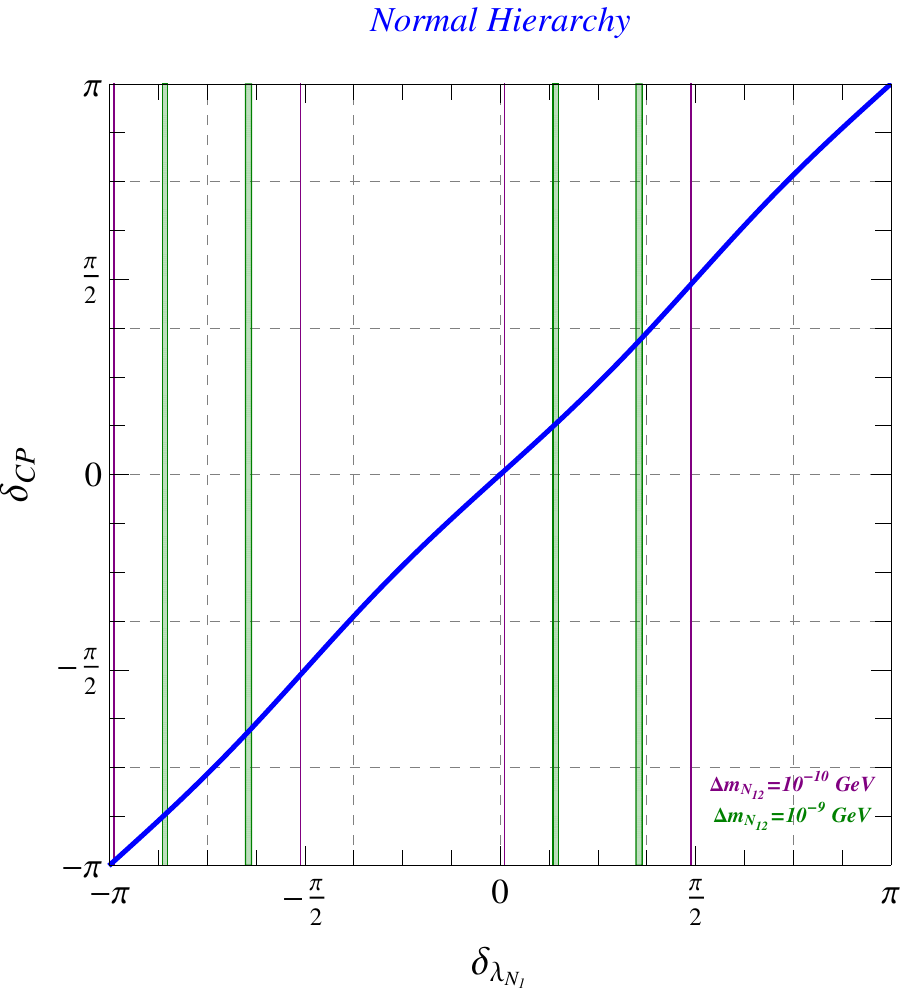}
    \end{minipage}\hspace{2 cm}
   \begin{minipage}{0.4\textwidth}
    \centering
    \includegraphics[scale=0.8]{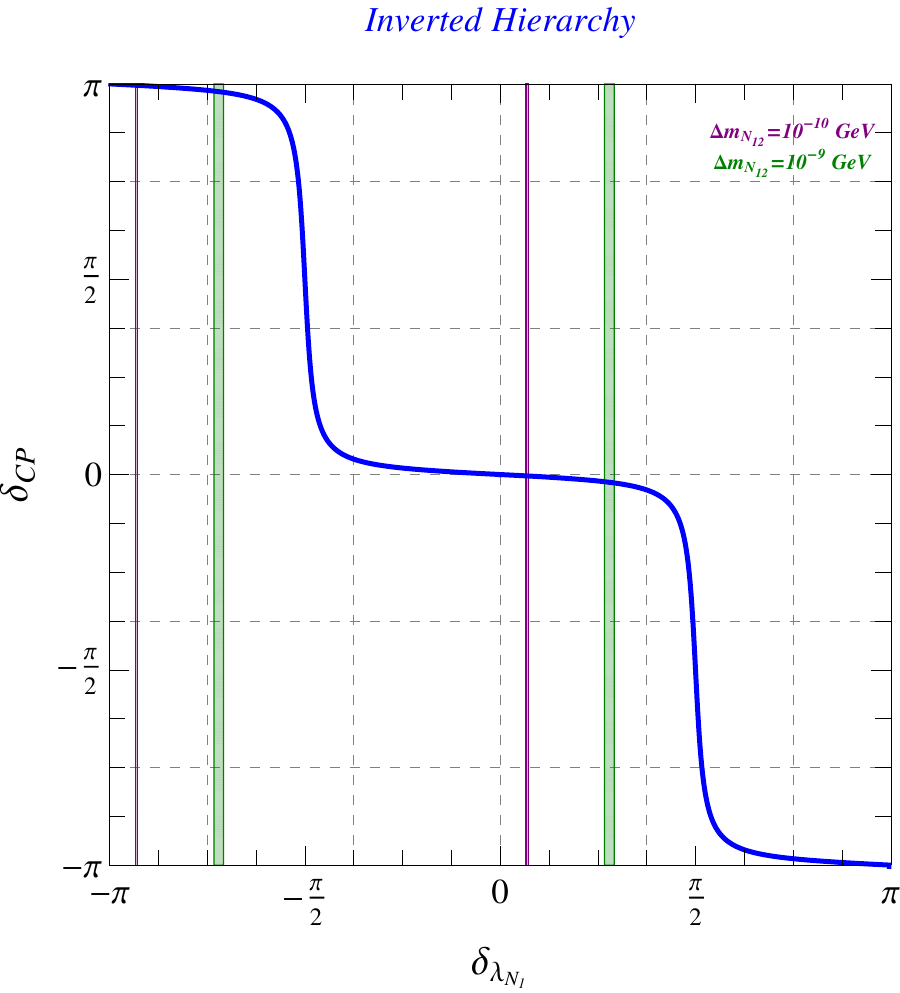}
    \end{minipage}
 \caption{\emph{ In the case of TeV $N_1$, confidence region in green~(purple) on  the phase of $\la_{N_1}$, $\de \la_{N_1}$, for $m_{N_2}-m_{N_1}
 \equiv \De m_{N_{12}}=10^{-9}$~($10^{-10}$) GeV.
 The blue line represents the correlation between
 $\de \la_{N_1}$ and $\de_{CP}$ in $U_{PMNS}$. }} \label{fig:DM_LEP_13}
\end{figure}

The fitting procedure is as follows.  The set of parameters involved are presented in Table~\ref{tab:bechmark_TeV} with $(N_a,N_b,N_c)=(N_3,N_2,N_1)$ and $\De m_{N_{12}}=10^{-9}$ and $10^{-10}$ GeV. Second, we vary $\la_{N_1}$ to find minimum $\chi^2$ taking into account all observables in Table~\ref{tab:observables}. Then, we fix $|\la_{N_1}|$ to the best-fit value and vary its phase $\de_{\la_{N_1}}$. The results are shown in Fig.~\ref{fig:DM_LEP_13}, 
where the green~(purple) band corresponds to the $99\%$ confidence region on $\de_{\la_{N_1}}$ for $\De m_{N_{12}}=10^{-9}$~$(10^{-10})$ GeV. We also display the corresponding $\de_{CP}$, ranging from $-\pi$ to $\pi$ as a function of $\de_{\la_{N_1}}$, with the blue line.\footnote{Again, $\th_{12}$, $\th_{23}$ and $\th_{13}$ are constrained to the first quadrant and Majorana phases range from $0$ to $\pi$ as discussed in Refs.~\cite{deGouvea:2008nm,deGouvea:2010iv}.} We have few observations based on Fig.~\ref{fig:DM_LEP_13}.
\begin{itemize}
\item For the NH case, $\de_{CP}$ is almost linearly proportional to $\de_{\la_{N_1}}$, which is not the case for the IH case. It is due to different chosen values for $m_{\nu_1}$ and $m_{\nu_3}$. In other words, if $m_{\nu_3}$ is zero, the linear proportionality will show up.

\item The lepton symmetry is actually proportional to the imaginary part of $\exp ( 2 i \de_{\la_{N_1}} )$ due to the fact $ Y_{\De L_\mu}$ and $ Y_{\De L_\tau}$ cancel each other because of the TBM pattern. Therefore, there will be no lepton asymmetry generated if $\de_{\la_{N_1}}=\pm\pi/2$. Furthermore, the physical range of $\de \la_{N_1}$ can be divided into four quadrants; two of them generate positive $\De Y_L$ while the other two produce needed negative $\De Y_L$ for the positive baryon asymmetry. 

\item The position of the confidence region depends on the maximal $\De Y_L$ at $\de \la_{N_1}=( \pi/4, - 3\pi/4)$. If the maximal value is much larger than the required $| \De Y_L | ~ (\sim 10^{-10})$, then the confidence region will be located near $0$, $\pm \pi/2$ and $ - \pi$ as in the NH case with $\De m_{N_{12}}=10^{-10}$ GeV. For the IH case with $\De m_{N_{12}}=10^{-9}$ GeV, the corresponding maximal value is quite close to $10^{-10}$, shifting the confidence to be around $\pi/4$ and $-3\pi/4$. 

\item With $\De m_{N_{12}}$ closer to $\Ga_{N_2}$~($\sim 10^{-11}$ GeV), the resonant enhancement becomes larger. That is why the confidence regions~(purple ones) with $\De m_{N_{12}}=10^{-10}$ GeV move toward to $0$, $\pm \pi/2$ and $ - \pi$ with respect to those of $\De m_{N_{12}}=10^{-9}$ GeV. For the IH case, $\de_{\la_{N_1}} \sim \pm \pi/2$ can not reproduce the correct $U_{PMNS}$ mixing angles and the mass-squared differences.    

\item It is quite interesting that the NH case has a confidence region near $\de_{CP}=-\pi/2$, which is preferred by the combined T2K and reactor measurements~\cite{Abe:2013hdq}.
\end{itemize}

To conclude, for TeV $N_1$, the simplest setup to reproduce observables in Table~\ref{tab:observables} in the presence of the flavor symmetry
is to involve only $\lee \de m_D \rii_{13}$, leading to the complex $R$ matrix such that low-mass leptogenesis can be achieved by the resonant enhancement, $m_{N_2} - m_{N_1} \sim \Ga_{N_2}$.

\section{S as inflaton} \label{sec:inflation}
In this section, we repeat the fitting procedure with $m_S$ and $m_{N_i}$ of $10^{13}$ GeV, a very different mass scale from the previous sections, in the context of the scalar $S$ being the inflaton $\phi$.  Recently, the BICEP2 experiment has reported a signal of inflationary gravitational waves in the $B$-mode power spectrum~\cite{Ade:2014xna}, which could be a hint of inflation.
We here explore the possibility of the scalar $S$ being the inflaton with the quadratic chaotic inflation~\cite{Linde:1983gd}.
We start with the summary of relevant equations on inflation. For recent reviews on inflation, see, for instance, Refs.~\cite{Kinney:2009vz,Baumann:2009ds,Martin:2013tda}.
In the limit of slow-roll inflation, the density (scalar) and tensor perturbations are related to the inflation potential $V(\phi)$ as:
\bea
\De_s^2 &\approx& \frac{1}{24\pi^2} \frac{V(\phi)}{M^4_{\mbox{\scriptsize{pl}}}} \frac{1}{\ep_V}, \nn\\
\De_t^2 &\approx&\frac{2}{3\pi^2} \frac{V(\phi)}{M^4_{\mbox{\scriptsize{pl}}}} ,
\eea
where $M_{\mbox{\scriptsize{pl}}}$ is the reduced Planck mass $(8\pi G)^{-1/2}~(=2.4\times 10^{18}$ GeV) and
\be
\ep_V= \left. \frac{M^2_{\mbox{\scriptsize{pl}}}}{2} \lee \frac{V^\prime}{V} \rii^2  \right\vert_{ \phi= \phi_{  \mbox{\scriptsize{cmb}}  }} 
= 2 \lee \frac{M_{\mbox{\scriptsize{pl}} } }{\phi_{ \mbox{\scriptsize{cmb}} }}  \rii ^2,
\ee
in which $V^\prime=dV/d\phi$ and $\phi_{ \mbox{\scriptsize{cmb}}}$ is the initial value of the inflaton field required to produce the observed Cosmic Microwave Background (CMB) fluctuations. $\phi_{ \mbox{\scriptsize{cmb}}}$ is related to $e$-folds $N_{ \mbox{\scriptsize{cmb}} }$ by
\be
\phi_{\mbox{\scriptsize{cmb}}} = 2 \sqrt{ N_{\mbox{\scriptsize{cmb}}} } M_{\mbox{\scriptsize{pl}}},
\ee
with $N_{\mbox{\scriptsize{cmb}}}\sim 40 -60$.
It implies $\phi_{cmb}$ will be super-Planckian and any flavor models based on effective theory approach will break down.
Therefore, when construction concrete UV-complete flavor models, one has to find a way
to highly suppress higher order terms like $\phi^4$ so that $m^2 \phi^2$ is dominant even with super-Planckian values for the inflaton.
Consequently, we have
\be
\ep_V= \frac{1}{2 N_{\mbox{\scriptsize{cmb}}} }.
\ee
 From the $Planck$ results~\cite{Ade:2013uln}, the scalar perturbation amplitude for $V=m^2_\phi \phi^2$ is $2.2\times 10^{-9}$,\,\footnote{In fact, many  inflation models have similar values of the scalar perturbation amplitude.} which in turns implies $m_\phi \sim 10^{13}$ GeV.
Furthermore, the tensor-to-scalar ratio $r$ is,
\be
r= \frac{ \De_t^2 }{ \De_s^2 } \approx 16 \ep_V = \frac{8}{N_{\mbox{\scriptsize{cmb}}}}\sim 0.16,
\ee
which is consistent with the BICEP2 results with  $r=0.20^{+0.07}_{-0.05}$ or $r=0.16^{+0.06}_{-0.05}$ after subtracting various dust models~\cite{Ade:2014xna}. In addition, the scalar spectral index, evaluated at CMB scales,
\be
n_s= 1 -\frac{2}{ N_{  \mbox{\scriptsize{cmb}} } } \sim 0.96,
\ee
which is also consistent with the $Planck$ results~\cite{Ade:2013uln}.

\begin{figure}[!htb!]
   \begin{minipage}{0.4\textwidth}
   \centering
   \includegraphics[scale=0.8]{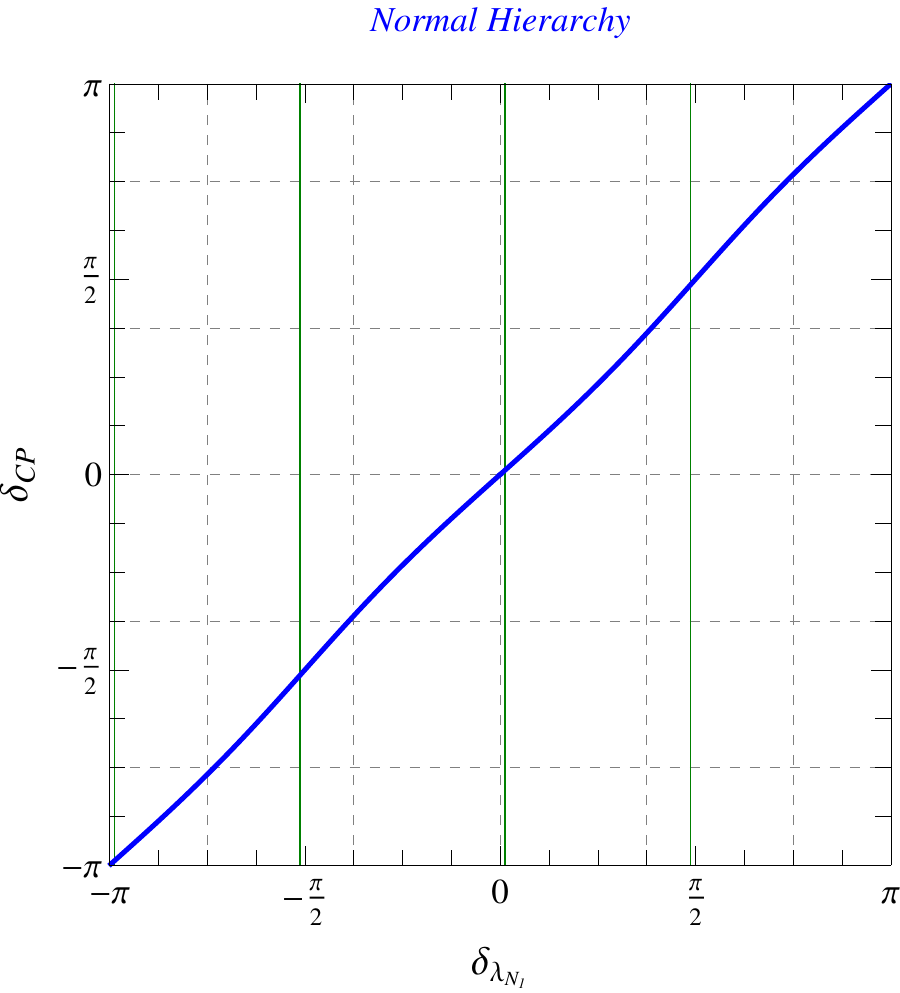}
    \end{minipage}\hspace{2 cm}
   \begin{minipage}{0.4\textwidth}
    \centering
    \includegraphics[scale=0.8]{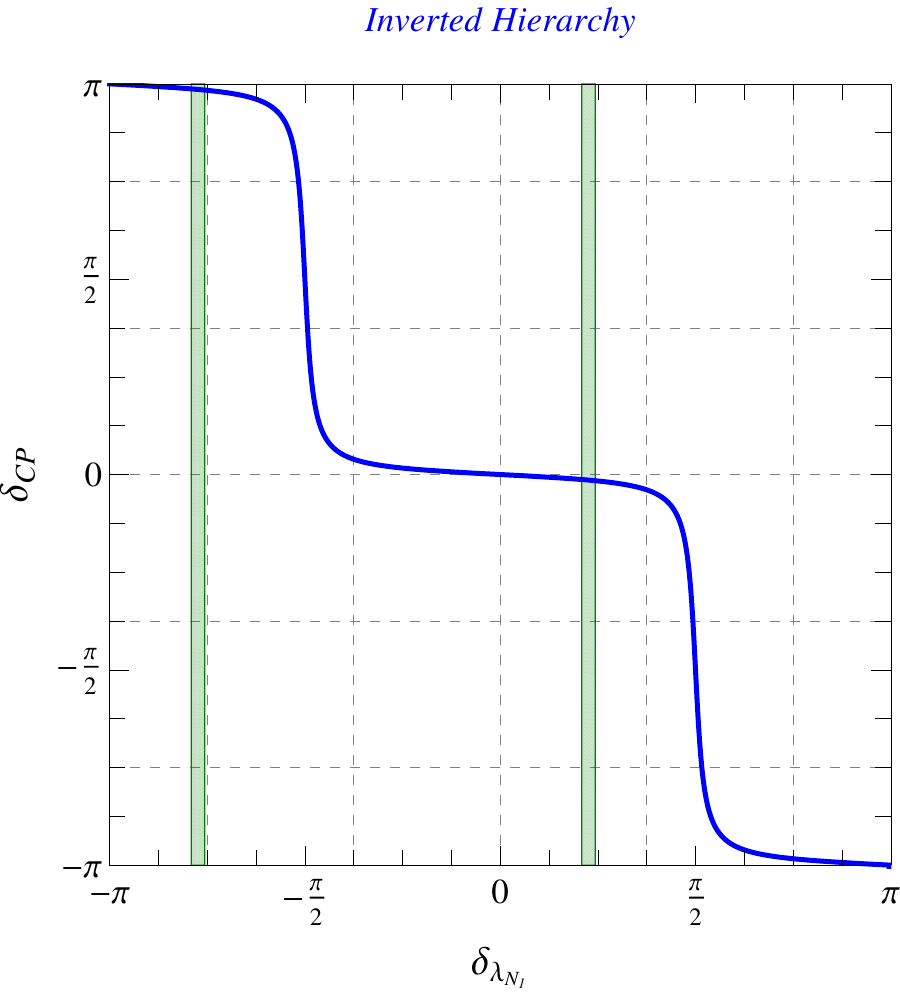}
    \end{minipage}
 \caption{\emph{
 Confidence region in green on $\de \la_{N_1}$ in the case of $S$ as the inflaton. The blue line represents the correlation between
 $\de \la_{N_1}$ and $\de_{CP}$.}}
 \label{fig:DM_LEP_13_inf}
\end{figure}

In the situation of $S$ being the inflaton with $m_S\sim 10^{13}$ GeV, to achieve leptogenesis and sizable $\th_{13}$, one has to make $m_{N_i} \gtrsim m_S$.   In this case, unlike the previous situation with TeV $N_1$, leptogenesis can be realized without resorting to the resonant enhancement since the heavy neutrinos satisfy the mass bounds, $m_{N_1} > 10^9$ GeV~\cite{Davidson:2002qv} and $m_{N_1} > 10^7$ GeV~\cite{Sierra:2013kba}. We have found that one needs only $(\de m_D)_{13}$ to reproduce the neutrino mixing angles and generate the correct lepton asymmetry.
 
\begin{table}[!h!]
\centering
\begin{tabular}{c c c c c c c c}
  \hline
  \hline
   & $m_{\nu_1}$ (eV) &  $ m_{\nu_2}$ (eV) &   $  m_{\nu_3}$ (eV) 
    & $\la_{N_a}$ & $\la_{N_b}$ &$\la_{\mu}$ &  $\la_{\tau}$ \\
  \hline 
 NH  &  0  & $8.66\times 10^{-3}$  & $4.89\times 10^{-2}$  & 0 &0 & 0 & 0\\
 \hline 
   IH &  $1.107 \times 10^{-1}$ & $1.11\times 10^{-1}$   & 0.1  & 0 &0 & 0 & 0 \\
  \hline\hline
       &$m_{N_1}$ (GeV) & $m_{N_2}$ (GeV) & $m_{N_3}$ (GeV) & $m_{S}$ (GeV) & $m_{\chi_1}$ (GeV) & $m_{\chi_2}$ (GeV) & $\la_e$  \\
  \hline
 NH/IH & $1.65\times 10^{13}$ & $3\times 10^{13}$ & $4.5\times 10^{13}$ & $1.5\times 10^{13}$ & 62 & 200 & 1    \\
  \hline
\end{tabular}

\caption{\emph{The Benchmark point for $m_N$'s and $m_S$ around the inflation scale, $10^{13}$ GeV. }}
\label{tab:bechmark_inf}
\end{table}

We adopt the same fitting produce as in Section~\ref{sec:DM13lepto}. The results are shown in Fig.~\ref{fig:DM_LEP_13_inf}, where the green band corresponds to the $99\%$ confidence region on $\de_{\la_{N_1}}$, and the blue line represents the correlation between $\de_{CP}$.
We briefly comment on the results, that are quite similar to those of TeV $N_1$.

\begin{itemize}

\item The behavior of the correlation between $\de_{CP}$ and $\de \la_{N_1}$ is the same as in the case of TeV $N_1$, i.e., determined by the value of $m_{\nu_1}$~($m_{n_3}$) for NH~(IH). Besides, the lepton symmetry is proportional to the imaginary part of $\exp ( 2 i \de_{\la_{N_1}} )$ because of the cancellation between $ Y_{\De L_\mu}$ and $ Y_{\De L_\tau}$ from the TBM pattern.

\item The lepton asymmetry comes from both the original vertex and wave function contribution (top panels of Fig.~\ref{fig:leptogenesis}); therefore, $m_{N_1}$ needs not to be close to $m_{N_2}$ as before.

\item Washout effects are not very efficient due to the fast expansion of the universe at such a high temperature so that the washout interactions can easily fall out of equilibrium. In addition, due to $m_{\chi_1} \ll m_{N_1}$, $\chi_1$ can carry the same lepton number as $L$ to a very good approximation. Hence,
we do not worry about the aforementioned washout interaction, $\chi_1 + S \leftrightarrow H^{\pm} + L^{\mp}$.  

\item The NH case also has a confidence region near $\de_{CP}=-\pi/2$ as above, favored by the combined T2K and reactor measurements~\cite{Abe:2013hdq}.

\end{itemize}

To conclude, for $m_S$ being the inflaton with mass of $10^{13}$ GeV, the single radiative correction $(\de m_D)_{13}$ to the Dirac mass matrix $m_D$ can render $\th_{13}\sim 9^\circ$ and
achieve leptogenesis. At the same time, one can have the correct DM density. It also ties the $CP$-violating phase $\de_{CP}$ in $U_{PMNS}$ with leptogenesis, that is absent from the original type-I seesaw.

\section{Conclusions} \label{sec:conclusion}

In the type-I seesaw, $\theta_{13}$ is zero if there exists an underlying residual $\mu-\tau$ symmetry. Furthermore, leptogenesis,
requiring the complex $R$ matrix~\cite{Casas:2001sr} characterizing heavy-light neutrino mixing, can not be achieved since a discrete flavor symmetry renders $R$ matrix real and diagonal.
Assuming the underlying residual flavor symmetry predicts the TBM neutrino mixing pattern, we here propose a simple toy model, where the additional particles, including the DM candidate, are introduced to break the residual flavor symmetry. 
Explicitly, an $SU(2)_L$ singlet fermion $\chi_1$, which is the DM candidate,   a fermonic $SU(2)_L$ doublet $\chi_2$ and a real gauge-singlet scalar $S$ generate the radiative corrections, $\de m_D$, to the neutrino Dirac mass matrix, leading to $\th_{13}\sim 9^\circ$ and providing both $CP$-violating and $CP$-conserving phases for leptogenesis. These additional particles are odd under the imposed $Z_2$ symmetry, which is used to guarantee the DM stability.

Keeping a spirit of minimality, we look for the minimum setup to achieve the aforementioned goals. We have found, for TeV right-handed neutrinos and sub-TeV $\chi$'s and $S$, one needs resonant leptogenesis, i.e., the mass difference between $N_1$ and $N_2$ is close to the decay width of $N_2$. Otherwise,
strong washout interactions lead to an insufficient lepton asymmetry. In this case, one requires only $(\de m_D)_{13}$ to simultaneously accommodate sizable $\th_{13}$ and leptogenesis, leading a connection between the $CP$ phase in the neutrino mixing matrix and leptogenesis. Interesting, with a small $N_1-N_2$ mass splitting in the NH case, the complex phase from the DM loop can generate $\de_{CP} \simeq -\pi/2$ favored by the T2K experiment~\cite{Abe:2013hdq}.  

On the other hand, in light of the recent BICEP2 results of the scalar-to-tensor ratio~\cite{Ade:2014xna}, $S$ being the inflaton, with the quadratic potential and the mass of $10^{13}$ GeV,
can explain the BICEP2 results very well. In this case, with the heavy neutrino mass of the same order, one also requires only $(\de m_D)_{13}$ to simultaneously accommodate sizable $\th_{13}$ and leptogenesis, because the corresponding Yukawa couplings are large enough to generate the lepton symmetry without the DM loop contribution or the resonance enhancement.  
Similarly, one of the confidence regions on $\de_{\la_{N_1}}$ in NH corresponds to $\de_{CP} \simeq -\pi/2$ preferred by the experiment.   

Finally, we would like to point out radiative corrections coming from particles outside of the dark sector could also render $\th_{13}$ nonzero.
It would, of course, spoil the connection between DM and neutrino physics advertised here. Therefore, one has to find a way to suppress or forbid
these kinds of corrections when building concrete models.

  \subsection*{Acknowledgments}

The author is especially thankful to Frank Deppisch, Valerie Domcke, Alfredo Urbano and Diego Aristizabal Sierra for many precious and enlightening discussions.
The author thanks Frank Deppisch, Jennifer Kile, Valerie Domcke and Alfredo Urbano for very useful comments on the draft, and John Ellis for pointing out the correlation between $\de_{CP}$ and leptogenesis.
The author thanks Chee Sheng Fong for indicating important washout effects and very helpful discussions.
The author is grateful for the hospitality of Academia Sinica (Taiwan), where this work was initiated. This work is supported by the London Centre for Terauniverse Studies (LCTS), using funding from the European Research Council via the Advanced Investigator Grant 267352.

\appendix

\section{Toy Model in $A_4$ } \label{sec:toyf}

In this Section, we construct a simple toy model in $A_4$, which mimics the model from Ref.~\cite{Chen:2012st} to demonstrate how the dark sector violates the residual flavor symmetry,
which leads to the TBM pattern.

The Lagrangian reads,
\be
\mathcal{L} \supset \mathcal{L}_1 + \mathcal{L}_2,
\ee
with 
\bea
\mathcal{L}_1 &=&  y L H N + \frac{\phi_E}{\Lambda} L \tilde{H} \lee y_e e^c  + y_\mu \mu^c + y_\tau \tau^c\rii 
+ m_N N N +  \ka \phi_N N N \nn\\
\mathcal{L}_2 &=&    \la   L D_2 S + \la_{H\chi} D_2 \tilde{H} D_1 + \la_N D_1 N S,
\eea
where $L=(L_e,L_\mu,L_\tau)$, $N=(N_1,N_2,N_3)$, $\phi_{N,E}=(\phi_{{N_1,E_1}}, \phi_{N_2,E_2},\phi_{N_3,E_3})$, 
$D_1=(\chi_1,sp_1,sp_2)$ and $D_2~(\tilde{D_2})=(\chi_2~(\tilde{\chi}_2),sp_3,sp_4)$ are triplets under $A_4$. Note that
we promote $\chi$s to $A_4$ triplets with the help of spurions~($sp$). On the other hand, one can also involve very massive physical fields into $D_i$ by playing with mass terms of $D_i$ such that the lightest mass eigenstate is $\chi_i$ and other massive particles have negligible contributions to the neutrino mass matrix. In any case, $A_4$ is broken by $\chi$s.
The particle content and corresponding quantum numbers are shown in Table.~\ref{tab:quantum_number_toyf}.

\begin{table}[!h!]
\centering
\begin{tabular}{c c c c c  c c c c c c  c c}
  \hline\hline
  Field & $L$ & $e^c$& $\mu^c$  & $\tau^c$ & $H$ & $N$ &  $D_1$ & $D_2$& $\tilde{D}_2$ & $S$ & $\phi_N$ & $\phi_E$ \\
  \hline
  $A_4 $ &  3 & 1 & $1^{\prime}$ & $1^{\prime\prime}$ & 1 & 3 & 3 & 3 & 3 & 1 &  3 & 3 \\

  \hline
 $SU(2)_L $ &  2 & 1 & 1 & 1 & 2 & 1 & 1& 2 & 2 & 1 & 1 & 1 \\
 \hline
  $U(1)_Y $ &  -1/2 & 1 & 1 & 1& 1/2 & 0 & 0 & 1/2 & -1/2 & 0 & 0 & 0 \\
 \hline
  $Z_2 $ &  + &  + & + & + & + & + & -- & -- & -- & -- & + & + \\
  \hline\hline
\end{tabular}
\caption{\emph{The particle content and corresponding quantum numbers in the toy model based on $A_4$. }}
\label{tab:quantum_number_toyf}
\end{table}
From $\mathcal{L}_1$, with $\lan \phi_E \ran \sim \lee v_E,0,0 \rii$, the charged lepton mass matrix is diagonal with masses proportional 
to $y_e$, $y_\mu$ and $y_\tau$, respectively.
The neutrino Dirac mass matrix are diagonal, $m_D = y \lan H \ran \mathbb{1}_{3\times 3}$ while the mass matrix for heavy neutrinos $N$
becomes,

\begin{align}
\label{eq:MN}
	M_N=
  \begin{pmatrix}
    \frac{2}{3} \kavu + m_N    &  -\frac{1}{3} \kavd  & -\frac{1}{3} \kavt \\
    -\frac{1}{3} \kavd &  \frac{2}{3} \kavt + m_N   &  -\frac{1}{3} \kavu \\
   -\frac{1}{3} \kavt &    -\frac{1}{3} \kavu  &   \frac{2}{3} \kavd + m_N  \\
  \end{pmatrix}.
\end{align}
The resulting light neutrino mass matrix is
\be
m_{\nu}= m_D M^{-1}_N m_D^T,
\ee
and it is easy to verify $m_{\nu}$ can be diagonalized by $U_{TBM}$ if $\lan \phi_{N_1}\ran=\lan \phi_{N_2}\ran=\lan \phi_{N_3}\ran$, i.e.,
\be
\hat{m}_{\nu}= U^T_{TBM} m_{\nu} U_{TBM},
\ee
where 
\be
U_{TBM}= \left(
  \begin{array}{c c c}
     \sqrt{ \frac{2}{3} }&  \frac{1}{\sqrt{3}}  & 0 \\
   - \frac{1}{\sqrt{6}} &  \frac{1}{\sqrt{3}}   &   \frac{1}{\sqrt{2}} \\
 -\frac{1}{\sqrt{6}}  & \frac{1}{\sqrt{3}}  &  -\frac{1}{\sqrt{2}}    \\
  \end{array}
\right).
\ee

Finally, radiative corrections coming from $D_i$ and S to $U_{TBM}$ will depend on how to embed $\chi_i$ into $D_i$.
Clearly, in the presence of spurions or heavy physical fields, the $A_4$ symmetry is violated. 
We would like to emphasize again that in this paper, we choose a model independent approach to study the
radiative corrections in a spirit of minimality to realize the nonzero $\th_{13}$, which could serve as a guiding principles
to build realistic models.
     
\section{DM relic density and DD } \label{sec:DMDD}

In the limit of small $\chi_1-\chi_2$ mixing angle $\theta$, and $\la_{H\chi}=\la_{H \tilde{\chi}}$, the annihilation cross section for  $s$-channel Higgs exchange is,

\bea
\lan \sigma v_{rel} \ran =&& \sin^2\theta \lee  \sum_f  \Theta(m_{\chi_1} -m_f)  \frac{ \la_{H\chi}^2 N_c m^2_f  m_{\chi_1}^2 v^2_{rel} }{4 \pi v^2}
\frac{ \lee 1 -  r_f \rii^{3/2} }{ \lee  4 m_{\chi_1}^2 -m_h^2  \rii^2 + m_h^2  \Ga_h^2 }  \nn \right. \\
&& + \Theta(m_{\chi_1} -m_W)  \frac{ \la_{H\chi}^2  m_{\chi_1}^4  v^2_{rel}}{8 \pi v^2}
\frac{ \lee 1 -  r_W^2 \rii^{1/2}  \lee 4- 4 r_W +3 r_W^2 \rii }{ \lee  4 m_{\chi_1}^2 -m_h^2  \rii^2 + m_h^2  \Ga_h^2 } \nn \\
  && \left. + \Theta(m_{\chi_1} -m_Z)  \frac{ \la_{H\chi}^2  m_{\chi_1}^4 v^2_{rel}}{16 \pi v^2}
\frac{ \lee 1 -  r_Z^2 \rii^{1/2}  \lee 4- 4 r_Z +3 r_Z^2 \rii }{ \lee  4 m_{\chi_1}^2 -m_h^2  \rii^2 + m_h^2  \Ga_h^2 }\rii,
\eea
 where $r_i=\lee m_i /m_{\chi_1} \rii^2$ for $i=(f,W,Z)$, $v_{rel}$ is the relative velocity, and $N_c$ is the color factor: 3 (1) for quarks (leptons).
 $m_f$, $m_W$ and $m_Z$ are the masses for final state fermions, W and Z boson, respectively. The step function $\Theta$ manifests the kinematical constraint. With $\lan \sigma v_{rel} \ran $, one can compute
 DM abundance including the thermal effect that is very important for the resonant enhancement.
 We refer readers to Ref.~\cite{Griest:1990kh} for more details.

The SI DM-nucleon cross-section via Higgs exchange is~\cite{Belanger:2008sj},

\be
\si_{SI}= c_{DM} \sin^2\th \frac{\mu_\chi^2}{\pi} \frac{( \la_{H\chi} M_N  f_N )^2}{ 2 m_h^4 v^2},
\ee
where $c_{DM}=1$ ($c_{DM}=4$) for Dirac (Majorana) DM, $M_N$ is the nucleon mass,
$\mu_\chi$ is the reduced DM-nucleon mass $\frac{m_{\chi_1} M_N }{ m_{\chi_1}+ M_N }$, $f_N=0.34$~\cite{Cline:2013gha},
and $v$ is the Higgs VEV($\sim 246$) GeV.

\section{ $\ep_{\al\al}$ in Leptogenesis computation } \label{sec:lep_com}
Here we present $\ep_{\al\al}$'s for three different situations: the original type-I
seesaw leptogenesis, the degenerate case $(m_{N_2}-m_{N_1} \sim \Ga_{N_2}  )$,
the DM loop contributions, respectively.

From Ref.~\cite{Covi:1996wh}, $\ep_{\al\al}$ in the original type-I seesaw leptogenesis, consisting of
the vertex and wave function contribution, is

\bea
\ep_{\al\al}=&& \frac{1}{8 \pi} \sum_{j  \neq 1  } \sum_{  \beta  } f( r_j ) \frac{ \mbox{Im}\left[ 
y^*_{\al j} y_{\al 1}  y^*_{\beta j} y_{\beta 1}  \right]   }{  \lee y^\dag y \rii_{11} +  |\la_{N_1}|^2 g_{kin}   } \nn\\
&&- \frac{1}{8 \pi}  \sum_{j  \neq 1  }  \frac{ m_{N_1} }{ m^2_{N_j} - m^2_{N_1} }
\frac{ \mbox{Im} \left\{  \left[  m_{N_j} \lee y^\dag y  \rii_{j1} +  m_{N_1} \lee y^\dag y  \rii_{1j}     \right]  
y^*_{\al j} y_{\al 1}   \right\}   }
{ \lee y^\dag y \rii_{11} +  |\la_{N_1}|^2 g_{kin}  },
\eea
where $r_j \equiv m^2_{N_j}/m^2_{N_1}$, $f(x)= \sqrt{x} ( 1 -(1+x)\ln[(1+x)/x ] ) $ and
\be
 g_{kin}= \frac{2 \lee m_{N_1}^2 -m_S^2 +m_{\chi_1}^2  \rii}{ m_{N_1}^3 }  \sqrt{ \frac{ \lee m_{N_1}^2 -m_S^2 +m_{\chi_1}^2 \rii^2 }  {4 m_{N_1}^2} - m_{\chi_1}^2 }.
 \ee
We here include the dilution from $N_1 \rightarrow \chi_1 S$, which does not generate the lepton asymmetry. 

In the limit of $N_1$ and $N_2$ being degenerate~$(m_{N_2}-m_{N_1} \sim \Ga_{N_2}  )$, where the lepton asymmetry is dominated
by the wave function contribution, we have~\cite{Pilaftsis:1997jf,Pilaftsis:2003gt}\footnote{Note that we have a different definition of Yukawa couplings from the Refs. }

\be
\ep_{res}=  \frac{ \mbox{Im} \left[   \lee y^\dag y \rii_{12}  \right]^2  }{ \lee \lee y^\dag y \rii_{11} +  |\la_{N_1}|^2 g_{kin} \rii   \lee y^\dag y \rii_{22}    } 
\frac{ \lee m^2_{N_2} - m^2_{N_1} \rii m_{N_1} \Ga_{N_2} }{ \lee m^2_{N_2} - m^2_{N_1} \rii^2  +  m^2_{N_1} \Ga^2_{N_2} },
\ee
where we have summed over all lepton flavors.

For the lepton asymmetry generated from the DM loop, $\ep_{\al\al}$ can be written as,
\be
\ep_{\al\al}= 2
\frac{ \mbox{Im} \lee  y^*_{\al 1}  \la^*_1 \la_{H\chi} \la_\al   \rii \mbox{Im} \lee f_{A_1}\rii
+  \mbox{Im} \lee  y^*_{\al 1}  \la_{N_1} \la_{H\chi} \la_\al   \rii \mbox{Im} \lee f_{A_2}\rii }
{  \lee y^\dag y \rii_{11}  +  |\la_{N_1}|^2 g_{kin}  },
\ee
where
\bea
f_{A_1}=&& \frac{ m_{N_1} m_{\chi_1}  }{ 16 \pi^2 \lee m_{N_1}^2-m_h^2 \rii }
 \lee  \lee m^2_S - m^2_{\chi_2} \rii
 C_0( m_h^2, 0,m^2_{N_1}, m_{\chi_1}^2, m_{\chi_2}^2,m^2_S  ) \right. \nn \\
 &&\left. +B_0(m^2_h, m_{\chi_1}^2, m_{\chi_2}^2) - B_0(m_{N_1}^2,m_{S}^2, m_{\chi_1}^2)   \rii,
 \eea
and
 \bea
f_{A_2}=&& \frac{ 1 }{ 16 \pi^2 \lee m_{N_1}^2-m_h^2 \rii }
 \lee  \lee m^2_S m^2_h - m^2_{N_1} m^2_{\chi_2} \rii
 C_0( m_h^2, 0,m^2_{N_1}, m_{\chi_1}^2, m_{\chi_2}^2,m^2_S  ) \right. \nn \\
 &&\left. + m^2_h  B_0(m^2_h, m_{\chi_1}^2, m_{\chi_2}^2) - m^2_{N_1}  B_0(m_{N_1}^2,m_{S}^2, m_{\chi_1}^2)   \rii.
 \eea
 $B_0$ and $C_0$ are Passarino-Veltman Integrals~\cite{Passarino:1978jh}.
 Note that  if $N_1$ decays before the electroweak phase transition, then the Higgs boson is massless, i.e., $m_h=0$.

\bibliography{DM_13}
\bibliographystyle{h-physrev}

\end{document}